\documentclass[review]{elsarticle}
\usepackage{makeidx}  

\makeatletter
\def\ps@headings{%
\def\@oddhead{\mbox{}\scriptsize\rightmark \hfil \thepage}%
\def\@evenhead{\scriptsize\thepage \hfil \leftmark\mbox{}}%
\def\@oddfoot{}%
\def\@evenfoot{}}
\makeatother
\usepackage{ wrapfig, color}
\usepackage{amssymb}
\usepackage{verbatim}
\usepackage{algorithm}
\usepackage{algpseudocode}
\usepackage{subfigure}
\usepackage{hyperref}

\usepackage{float}
\usepackage{graphicx}
\usepackage{epstopdf}
\usepackage{multirow}
\usepackage{amsmath}
\usepackage[nomargin,inline,marginclue,draft]{fixme}

\usepackage{mathtools}

\makeatletter
\renewcommand*\FXLayoutInline[3]{%
  {\@fxuseface{inline}\ignorespaces[#3 \fxnotename{#1}: #2]}}
\makeatother

\begin{document}

\begin{frontmatter}

\title{A Heuristic Approach to Protocol Tuning\\ for High Performance Data Transfers}

\author[unr]{Engin~Arslan\corref{mycorrespondingauthor}}
\address[unr]{University of Nevada, Reno, 1664 N Virginia St, Reno, NV 89557}
\cortext[mycorrespondingauthor]{Corresponding author}
 \ead{earslan@unr.edu}

\author[ub]{Tevfik~Kosar}
\address[ub]{University at Buffalo, SUNY, 338 Davis Hall, Buffalo, New York 14260}

\begin{abstract}
Obtaining optimal data transfer performance is of utmost importance to today's data-intensive distributed applications and wide-area data replication services.
Doing so necessitates effectively utilizing available network bandwidth and resources, yet in practice transfers seldom reach the levels of utilization they potentially could.
Tuning protocol parameters such as pipelining, parallelism, and concurrency can significantly increase utilization and performance,
however determining the best settings for these parameters is a difficult problem, as network conditions can vary greatly between sites and over time.
Nevertheless, it is an important problem, since poor tuning can cause either under- or over-utilization of network resources and thus degrade transfer performance.
In this paper, we present three algorithms for application-level tuning of different protocol parameters for maximizing transfer throughput in wide-area networks. 
Our algorithms dynamically tune the number of parallel data streams per file (for large file optimization), the level of control channel pipelining (for small file optimization), 
and the number of concurrent file transfers to increase I/O throughput (a technique useful for all types of files).
The proposed heuristic algorithms improve the transfer throughput up to 10x compared to the baseline and 7x compared to the state of the art solutions.


\end{abstract}
\end{frontmatter}

\section{Introduction}
\label{sec:intro}
Despite the increasing availability of high-speed wide-area networks and
the use of modern data transfer protocols designed for high performance,
file transfers in practice attain only fractions of theoretical
maximum throughputs, leaving networks underutilized and users unsatisfied.
This fact is due to a number of confounding factors, such as
under-utilization of end-system CPU cores, low disk I/O speeds, server
implementations not taking advantage of parallel I/O opportunities,
background traffic at inter-system routing nodes, and unsuitable system-level tuning
of networking protocols.

The effects of some of these factors can be mitigated to varying degrees
through the use of techniques such as command pipelining, transport-layer
parallelism, and concurrent transfers using multiple data channels.
The degree to which these techniques are utilized, however, has the
potential to negatively impact the performance of the transfer and the
network as a whole. Too little use of one technique, and the network
might be underutilized; too much, and the network might be overburdened to
the detriment of the transfer and other users. Furthermore, the optimal
level of usage for each technique varies depending on the network and end-system conditions, 
meaning no single parameter combination is optimal for all different scenarios. 

We propose dynamic optimization algorithms to determine which combination of parameters is ``just right" for a given transfer task. Main contributions of this paper are: (i) optimization of dataset clustering for heterogeneous datasets that contain small and large files together, (ii) a heuristic approach to estimate parameter values to be used in transfer, and (iii) three novel scheduling algorithms to improve data transfer throughput. We have run extensive experiments in wide- and local-area networks and using real and synthetic datasets. The experimental results are very promising, and our algorithms outperform other existing solutions in this area.


The remainder of this paper is organized as follows. The next section presents the related 
work. Section~\ref{sec:algorithms} introduces the dynamic protocol tuning 
algorithms we propose. Section~\ref{sec:results} presents the performance evaluation results of 
our algorithms. Section~\ref{sec:conclusion} concludes the paper with our discussion and future directions.

\section{Related Work}
\label{sec:related}
Liu et al.~\cite{R_Liu10} developed a tool which optimizes multi-file transfers by opening multiple GridFTP~\cite{NDM_2012} threads.
The tool increases the number of concurrent flows up to the point where transfer performance degrades.
Their work only focuses on concurrent file transfers, and other transfer parameters are not considered.
%
Globus Online~\cite{globusonline}
sets the pipelining, parallelism, and concurrency parameters to specific values for three different file sizes (i.e., less than 50MB, larger than 250MB, and in between).
However, the protocol tuning Globus Online performs is non-adaptive; it does not change depending on network conditions and transfer performance.
%
Similar Managed File Transfer (MFT) systems were proposed which used a subset of these parameters in an effort to improve the end-to-end data transfer throughput~\cite{WORLDS_2004, ScienceCloud_2013, Royal_2011, IGI_2012}.

Other approaches aim to improve throughput by opening flows over multiple paths between end-systems~\cite{Raiciu:2010:DCN:1868447.1868457,Khanna:2008:UOE:1413370.1413418},
however there are cases where individual data flows fail to achieve optimal throughput because of end-system bottlenecks.
Several others propose solutions that improve utilization of a single path by means of parallel streams~\cite{Altman06paralleltcp,Hacker:2005:ADB:1203492.1318153,R_Dinda05},
pipelining~\cite{TCP_Pipeline, farkas2002, R_Bres07},
and concurrent transfers~\cite{kosar04, Kosar09, R_Liu10}.
Although using parallelism, pipelining, and concurrency may improve throughput in certain cases, an optimization algorithm
should also consider system configuration, since end systems may present factors (e.g., low disk I/O speeds or over-tasked CPUs) which can introduce bottlenecks.

In our previous work~\cite{R_Yildirim11a}, we proposed network-aware transfer optimization by automatically detecting bottlenecks and improving
throughput via utilization of network and end-system parallelism.
%
We developed three highly-accurate models~\cite{R_Yin11, R_Yildirim11, DISCS12} which would require as few as three sampling points to provide accurate predictions for the optimal parallel stream number.
These models have proved to be more accurate than existing similar models \cite{R_Hacker02, R_Dinda05} which lack in predicting the parallel stream number that gives the peak throughput.
We have developed algorithms to determine the best sampling size and the best sampling points for data transfers by using bandwidth, Round-Trip Time (RTT), or Bandwidth-Delay Product (BDP) \cite{ccsa12}. 
We have analyzed the combined effect of these transfer parameters on end-to-end data transfer throughput, and developed several predictive (offline) and dynamic (online) algorithms to choose the best parameter combination to minimize the delivery time of the data~\cite{tcc16, europar13, zulkar-ndm14, sc16}.

\section{Dynamic Protocol Tuning Algorithms}
\label{sec:algorithms}
Different transfer parameters such as pipelining, parallelism, and concurrency play a significant role in affecting achievable transfer throughput.
However, setting the optimal levels for these parameters is a challenging problem,
and poorly-tuned parameters can either cause underutilization of the network or overburden the network and degrade the performance due to increased packet loss,
end-system overhead, and other factors.

Among these parameters, {\bf pipelining} specifically targets the problem of transferring a large numbers of small files~\cite{TCP_Pipeline, farkas2002, Cluster_2015}.
In most control channel-based transfer protocols, an entire transfer must complete and be acknowledged before the next transfer command is sent by the client.
This may cause a delay of more than one RTT between individual transfers.
With pipelining, multiple transfer commands can be queued up at the server, greatly reducing the delay between transfer completion and the receipt of the next command.
{\bf Parallelism} sends different portions of the same file over parallel data streams (typically TCP connections),
and can achieve high throughput by aggregating multiple streams~\cite{R_Hacker05, DADC_2008, DADC_2009}.
{\bf Concurrency} refers to sending multiple files simultaneously through the network using different data channels at the same time, and is especially useful for increasing I/O concurrency in parallel disk systems~\cite{R_Liu10, Thesis_2005, JGrid_2012}.

\begin{figure*}[!h]
\begin{centering}
\subfigure[Pipelining]{
\includegraphics[keepaspectratio=true,angle=0,width=59mm]{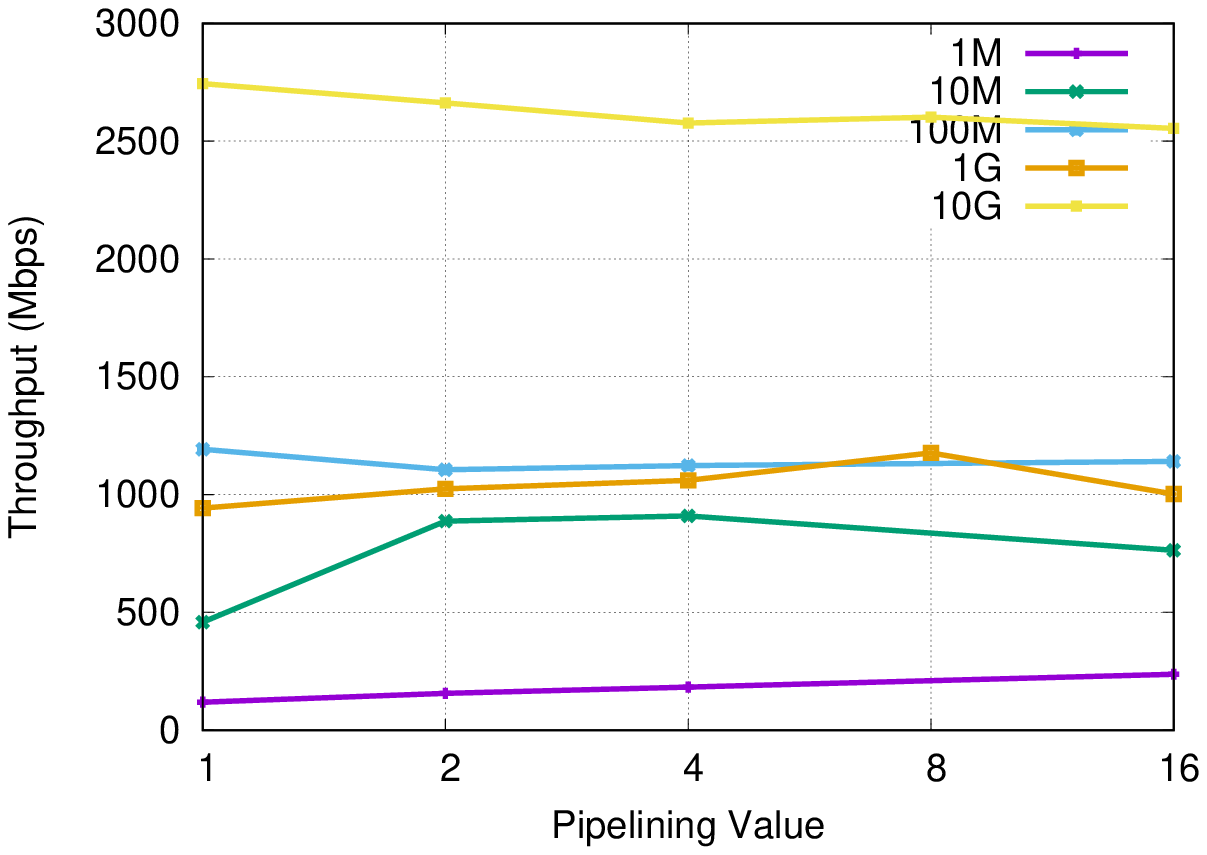}
\label{fig:xsede_ppq}}
\hspace{-4mm}
\subfigure[Parallelism w/ fixed pipelining]{
\includegraphics[keepaspectratio=true,angle=0,width=59mm]{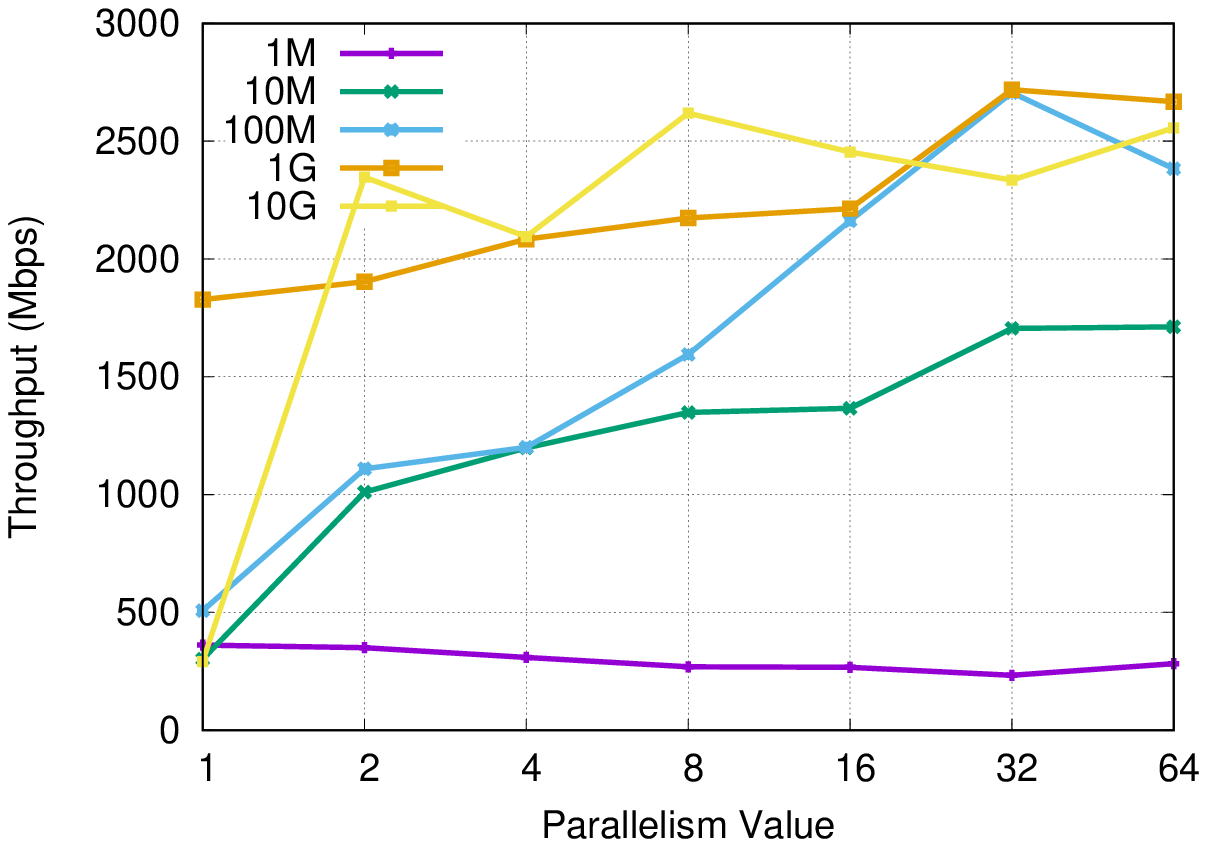} 
\label{fig:xsede_ppq+p}}
\hspace{-4mm}
\subfigure[Concurrency w/ fixed pipelining,parallelism]{
\includegraphics[keepaspectratio=true,angle=0,width=59mm]{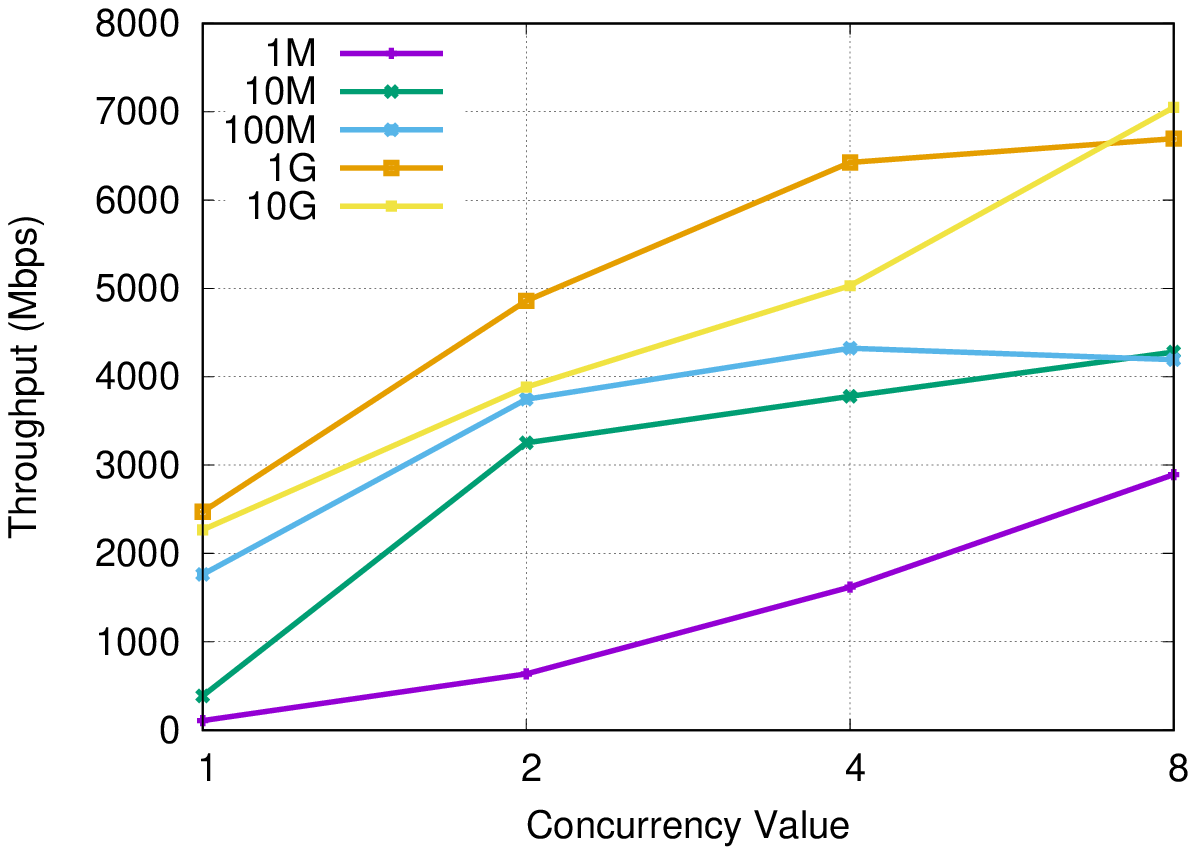}
\label{fig:xsede_ppq+p+cc}}
\caption{Effect of protocol parameters on transfer throughput for different file sizes in XSEDE network.} 
\label{fig:param-effect-xsede}
\end{centering}
\end{figure*}

\begin{figure*}[!h]
\begin{centering}
\subfigure[Pipelining]{
\includegraphics[keepaspectratio=true,angle=0,width=59mm]{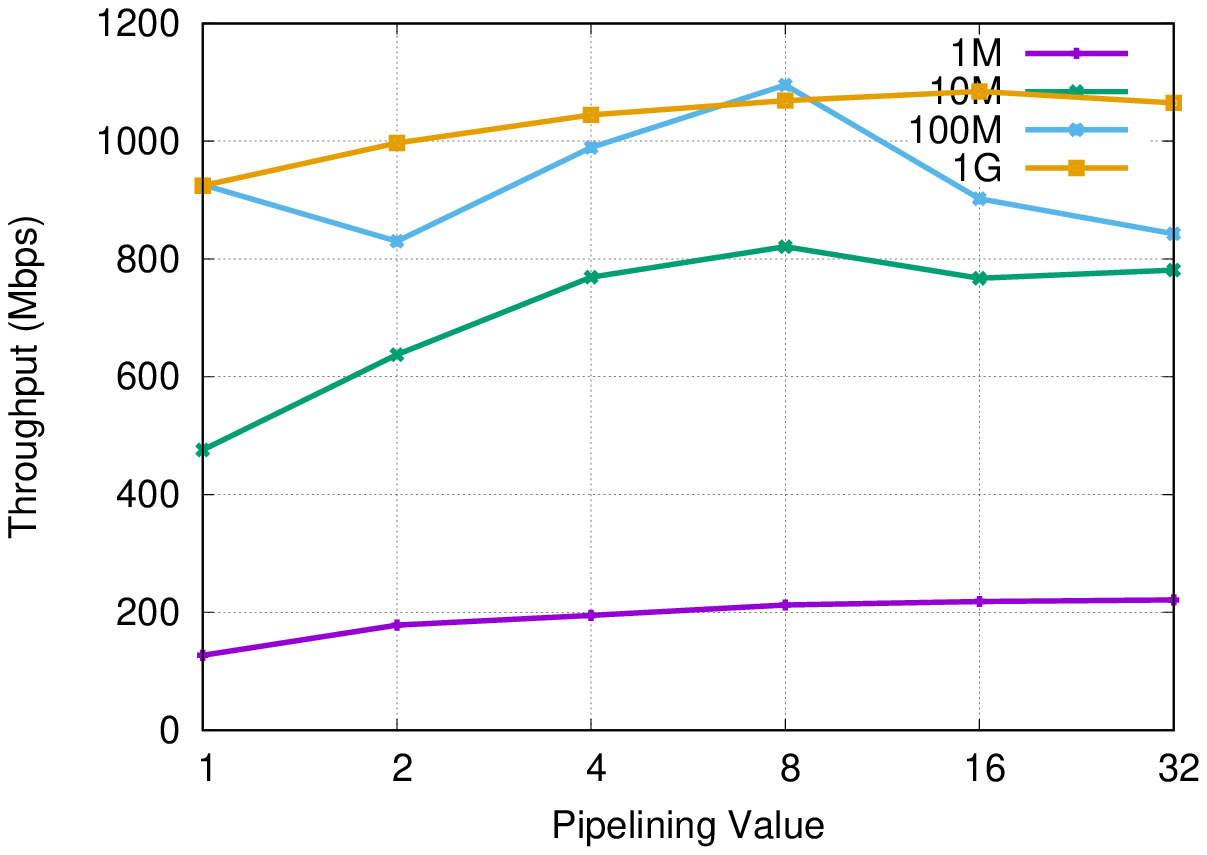}
\label{fig:loni_ppq}}
\hspace{-4mm}
\subfigure[Parallelism w/ fixed pipelining]{
  \includegraphics[keepaspectratio=true,angle=0,width=59mm]{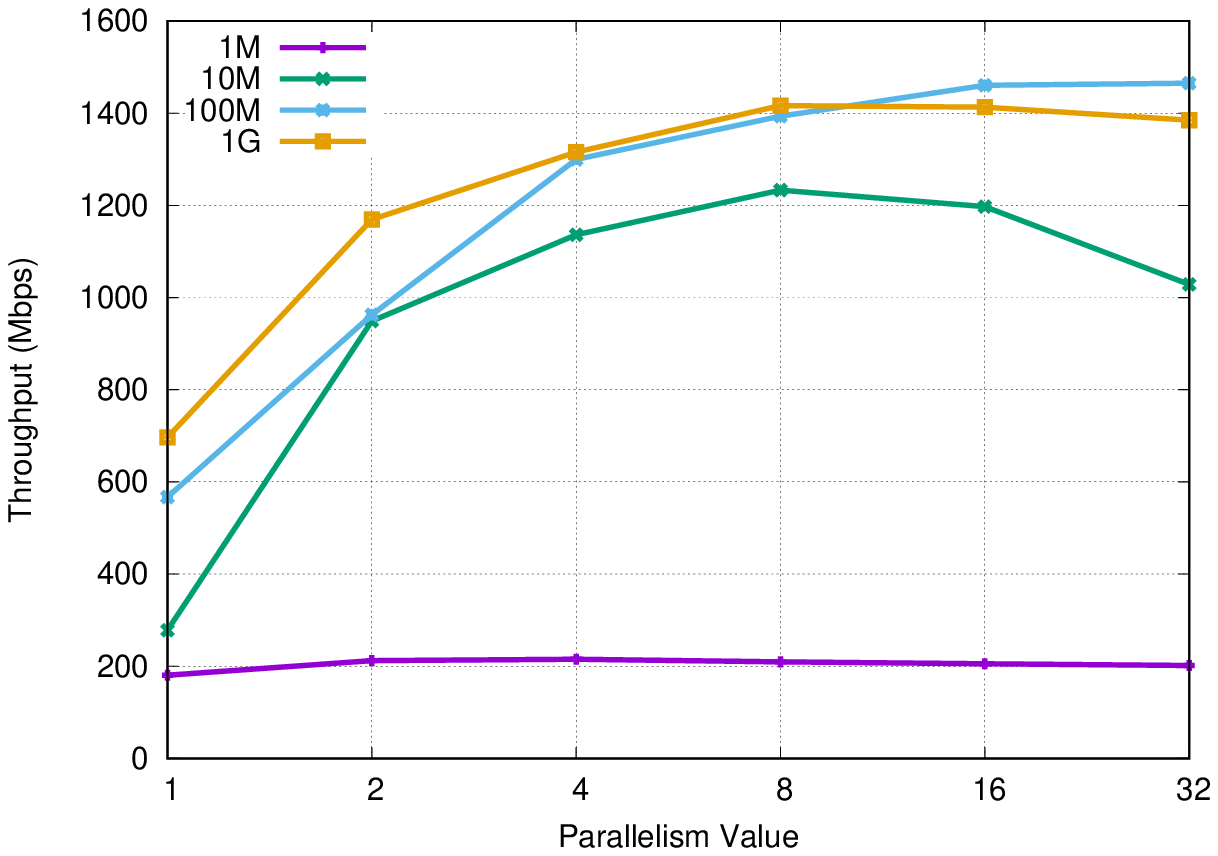} 
\label{fig:loni_ppq+p}}
\hspace{-4mm}
\subfigure[Concurrency w/ fixed pipelining,parallelism]{
   \includegraphics[keepaspectratio=true,angle=0,width=59mm]{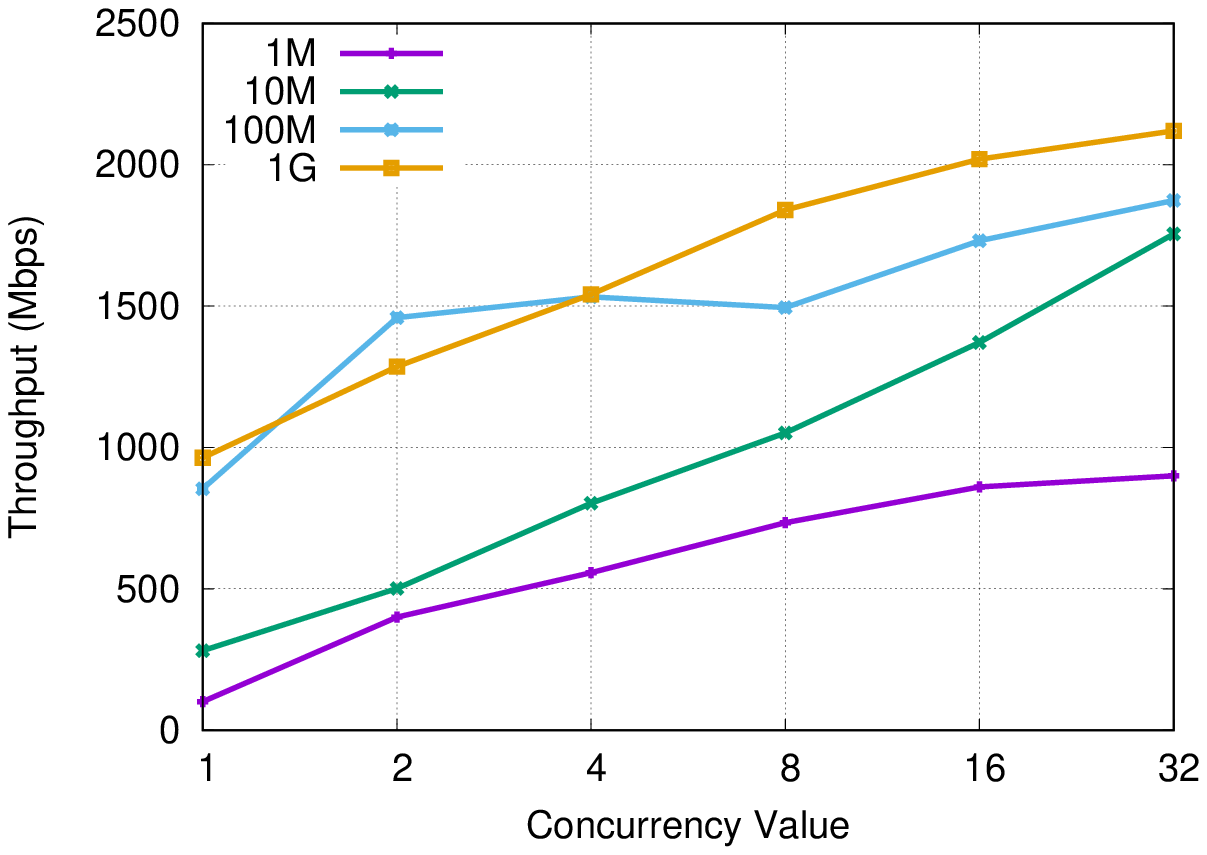}
\label{fig:loni_ppq+p+cc}} 
\caption{Effect of protocol parameters on transfer throughput for different file sizes in LONI network.} \label{fig:param-effect-loni}
\end{centering}
\end{figure*}

To analyze the effects of different parameters on the transfer of different file sizes,
we initially conducted experiments for each of the parameters separately, as shown in Figures~\ref{fig:param-effect-xsede} and~\ref{fig:param-effect-loni}. We run our experiments on XSEDE~\cite{xsede} and LONI~\cite{LONI} production-level high-bandwidth networks.
Although both of the networks have 10G network bandwidth between sites, XSEDE provides higher throughput in end-to-end (disk-to-disk) transfers despite the high RTT between its sites.
This is mainly due to the highly tuned and parallelized disk sub-systems at the XSEDE sites. 

We generated five datasets for five different file sizes and transferred each dataset, only changing one parameter (i.e., pipelining, parallelism or concurrency) at a time to observe the individual effect of each parameter. Then we introduced other parameters one by one. Figures~\ref{fig:xsede_ppq} and~\ref{fig:loni_ppq} show that pipelining can increase the throughput of small files by up to 2x while its impact becomes negligible for large files. On the contrary, parallelism helps to improve transfer throughput for large files significantly while it has no impact (if not negative) on small files as shown in Figures~\ref{fig:xsede_ppq+p} and~\ref{fig:loni_ppq+p}. Thus, it is necessary to use different parameter values for different file sizes. This requires separating small and large files if they are mixed in a dataset. Moreover, we observed that some of the datasets exhibit similar behavior to changing parameter values. For example, pipelining has either limited or negative impact on 1G and 10G datasets as shown in Figure~\ref{fig:param-effect-xsede}. Similarly, the throughput for both datasets increases with increased parallelism and concurrency levels. Hence, we analyzed the impact of creating different number of chunks (aka partitions\footnote{In the rest of the paper, the terms  chunk, file group, and partition are used interchangeably, and they refer to a set of files. Each file is treated as single unit and not partitioned into smaller files.}) from a given mixed dataset in the Section~\ref{sec:results}.

Concurrency is the most broadly effective parameter for all file sizes in both networks as it helps to improve disk I/O throughput by means of reading/writing multiple files simultaneously. While it is the most effective parameter, it incurs the most overhead to the end systems at the same time by increasing CPU usage at the end systems~\cite{ismail-suscom,ismail-sc15}. Hence, we use the concurrency level as the pivot parameter in the comparison of different algorithms in this section.

\begin{table}[!h]
\begin{centering}
\resizebox{\textwidth/2}{!}{\begin{tabular}{ |c| c| c|}
\hline
 \multirow{2}{*}{\bf Specs}&\multicolumn{1} {|c|}{\bf XSEDE}& \multicolumn{1}{|c|} {\bf LONI} \\
&Lonestar-Gordon&Quenbee-Painter\\
Bandwidth (Gbps)& 10 & 10  \\
\hline
RTT (ms)& 60  & 10   \\
\hline
Buffer Size (MB)& 32 & 16   \\
\hline
BDP (MB)& 75  & 9   \\
\hline
\end{tabular}}
\caption{System specifications of test environments} \label{tab:system-spec}
\end{centering}
\end{table}

Based on the results we observed in XSEDE and LONI networks, we devise heuristic equations to estimate the parameter values for different file sizes. We propose three dynamic protocol tuning algorithms to schedule the transfer of chunks using estimated parameter values:
(i) the ``Single-Chunk (SC)" algorithm, which separates files into chunks based on file size, and then transfers each chunk with its optimal parameters;
(ii) the ``Multi-Chunk (MC)" algorithm, which likewise creates chunks based on the file size, but rather than scheduling each chunk separately, it co-schedules and runs small-file chunks and large-file chunks together in order to balance and minimize the effect of poor performance of small file transfers;
(iii) the ``Pro-Active Multi-Chunk (ProMC)" algorithm, which instead of allocating channels equally among chunks, considers chunk size and type, and improves the performance especially if the small files dominate the dataset. Although proposed algorithms differ in terms of how to schedule chunks, they share methods that determine chunks and parameter values for chunks which are explained in the next section.

\begin{figure}[!h]
\begin{centering}
\begin{tabular}{c}
 \includegraphics[keepaspectratio=true,angle=0,width=100mm]{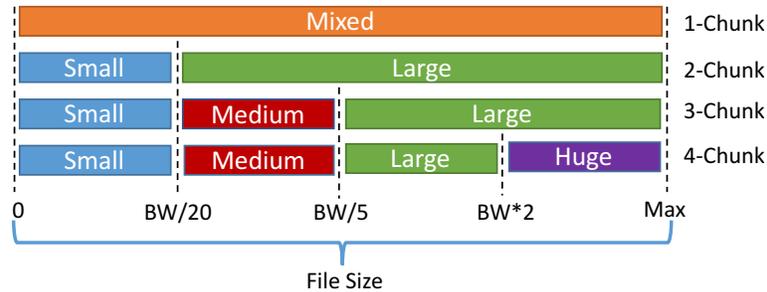}
\end{tabular}
\caption{Thresholds to partition datasets into chunks (BW = network bandwidth).} \label{fig:chunk-count}
\end{centering}
\end{figure}

{

\subsection{Heuristic Approach to Creating Chunks and Estimating Parameter Values}

Files with different sizes need different sets of transfer parameters to obtain optimal throughput. As an example, pipelining and data channel caching would mostly improve the performance of small file transfers, whereas parallel streams would be beneficial if the files are large. Optimal concurrency levels for different file sizes would be different as well. Hence all three algorithms starts with separating files into chunks based on file size. 

Several methods have been proposed to partition a mixed dataset into subgroups based on file size. Yildirim et al.~\cite{tcc16} creates up to five chunks when a mixed dataset is given and uses Bandwidth Delay Product (BDP) to determine cut off points. In our earlier works~\cite{europar13, sc16}, we create up to four chunks again using BDP. On the other hand, Globus Online~\cite{globusonline} and {\it{globus-url-copy}} (command line tool for running GridFTP transfers) treat whole dataset as single chunk (aka 1-chunk). While {\it{globus-url-copy}} requires manual parameter tuning, Globus Online uses different parameter values based on average file sizes of dataset. Datasets whose average file size is smaller than 50 MB are considered as small, between 50 MB and 250 MB as medium and larger than 250 MB as large files. To evaluate the effect of the number of chunks, our proposed algorithms takes the number of chunks as input and creates up to four chunks as shown in Figure~\ref{fig:chunk-count}. For example, if the number of chunks is specified as 3, then BW/20 and BW/5 will be used as thresholds (where BW is the network bendwidth) and up to three chunks will be created if there are enough files in the dataset. While it is possible that different cutoff points may return different results, the results in Section~\ref{sec:results} show that creating more than two chunks has little impact on transfer throughput so we do not focus on finding optimal cutoff values in this work.

Once chunks are created, we calculate the optimal parameter combination for each chunk using heuristic equations as given in Algorithm~\ref{alg:parameterEstimation}. Pipelining and concurrency are the most effective parameters for small files transfers in overcoming poor network utilization, so it is especially important to choose the best pipelining and concurrency values for such transfers. We set the pipelining values by considering the BDP and average file size of each chunk (line 3) such that it will return large values for small files. As the average file size of a chunk increases, it returns smaller pipelining value since it does not further improve performance, and can even cause performance degradation by causing load imbalance between channels. For parallelism, we consider BDP, average file size, and the TCP buffer size (line 4). Since parallelism is especially helpful for large files when maximum TCP buffer size is smaller than BDP, we use two inputs in determining the parallelism level. $\frac{BDP}{bufferSize}$ checks if buffer size is smaller than BDP and if so it will return the number of required parallel streams to overcome the buffer size limitation. $\frac{avgFileSize}{buferSize}$ controls whether or not the file is big enough to benefit from multiple streams. Even though buffer size limitation exists, small files will not be able to fully utilize available buffer thus there is no real benefit of using multiple streams to transfer small files. Finally, the concurrency level is determined by three inputs which are BDP, average file size, and user input of maximum concurrency. Since small files obtain much smaller throughput than large files for the same concurrency value (which can be observed in concurrency level 1 of Figure~\ref{fig:xsede_ppq+p+cc} and~\ref{fig:loni_ppq+p+cc}), small file types require large concurrency values to achieve high transfer throughput. Hence, $\frac{BDP}{avgFileSize}$ will return larger values for smaller chunks. While concurrency is the most powerful parameter for both small and large file types, it incurs the most cost by opening several processes for data and communication channels at the end servers. Thus, it is advised by system administrators to use it judiciously. $maxCC$ allows user to determine the maximum number of concurrent file transfers to use for a given transfer task such that even if $\frac{BDP}{avgFileSize}$ returns very large values, it will define an upper limit. Finally, we set lower limit for concurrency as 2 since concurrency is mostly helpful and $\frac{BDP}{avgFileSize}$ may return less than one if chunk's average file size is larger than BDP.

\subsection{Single-Chunk (SC) Algorithm}\label{sec:sc}

Single-Chunk (SC) aims to fine-tune transfer parameters for different file sizes using a divide and transfer approach. It first categorizes files into file chunks based on file size as described in previous section, then transfers them one by one. For example, if a given dataset consists of small and large files, SC will separate them into chunks, then will determine parameter values to be used for each chunk using Algorithm~\ref{alg:parameterEstimation}. Then, chunks will be transferred sequentially with their own parameter values.

Since Algorithm~\ref{alg:parameterEstimation} returns different parallelism values for different chunk types, the same data connection may not be reused for different chunk transfers. The parallelism value can only be set once at the time of connection establishment and cannot be altered afterwards. Thus, although creating more chunks allows to set more accurate parameters for each file types, it may cause a delay due to connection setup/tear down overhead.

\begin{algorithm}[!htb]
\scriptsize
\centering
\caption{--- Calculation of protocol parameter values} \label{alg:parameterEstimation}
\begin{algorithmic}[1]
  \Function{findOptimalParameters}{avgFileSize, BDP, bufferSize, maxCC}
	\State $pipelining = \frac{BDP}{avgFileSize}$
	\State $parallelism = Min (\left\lceil\frac{BDP}{bufferSize}\right\rceil, \left\lceil\frac{avgFileSize}{bufferSize}\right\rceil)$
	\State $concurrency = Min(Max(\frac{BDP}{avgFileSize}, 2), maxCC)$\label{alg:cc}
	\State \Return (pipelining, parallelism, concurrency)
\EndFunction
\end{algorithmic}
\end{algorithm}

\begin{algorithm}[t]
\scriptsize
\centering
\caption{--- Multi-Chunk (MC) Scheduling} \label{alg:mc}
\begin{algorithmic}[1]
\Statex
\Function{transfer}{source, destination, BW, RTT, maxCC} \label{mc:input}
	\State $BDP = BW*RTT$
	\State $allFiles = fetchFileListFromSource()$
	\State $chunks = partitionFiles(allFiles,BDP)$
	\For{$i=0; i<chunks.length; i++$}
		\State $chunks[i].optParams = findOptimalParameters(chunk)$
	\EndFor
	\While{$maxCC > 0$}
		\State $Chunk~c = nextChunk()$
		\Comment{Round-robin from set of \{Huge,Small,Large,Middle\}} \label{mc:rr}
		\State $c.concurrency += 1$ \label{mc:concurrencyLevel}
		\Comment{Add channel  to the chunk}
		\State $maxCC-= 1$
	\EndWhile
	\For{$i=0; i<chunks.length; i++$}
		\State $transfer(chunks[i])$\label{mc:scheduleall}
		\Comment{Run in the background}
	\EndFor
\EndFunction
\end{algorithmic}
\end{algorithm}

\subsection{Multi-Chunk (MC) Scheduling}\label{sec:mc}
Multi-Chunk (MC) algorithm focuses mainly on minimizing the effect of small chunks on the overall transfer throughput for mixed datasets.
Based on the results obtained from the SC approach, we deduced that even after choosing the best parameter combination for each chunk, throughput values obtained during the transfer of the small files (e.g. Small and Medium chunks) are significantly lower compared to that of large files (e.g. Large and Huge chunks) due to the high overhead of reading too many files from disk and under-utilization of the network pipe.
Depending on the weight of small files over total dataset size, overall throughput can be much less than the throughput of large chunk transfers.
Thus, we developed the MC algorithm which aims to minimize the effect of poor transfer throughput of a dataset dominated by small files. Similar to SC, MC uses {\it{findOptimalParameters}} method to calculate the values of parallelism and pipelining, however it defines concurrency value to be equal to $maxCC$ to run as many chunks simultaneously as possible.

After chunks are created, MC distributes available concurrency (the number of concurrent file transfer channels) among chunks using round-robin on set of \{Huge, Small, Large, Medium} (line~\ref{mc:rr} in Algorithm~\ref{alg:mc}).
The ordering of chunks provides better chunk distribution if the number of channels is less than the number of chunks. As an example, if $maxCC$ is given as 8 and three chunks (Small, Medium, and Large) are created from dataset, then concurrency distribution among chunks will be (3,2,3) for (Small, Medium, and Large) set. After channel distribution is completed, the MC schedules chunks concurrently using the calculated concurrency level for each chunk (line~\ref{mc:scheduleall}). That is, for the above example, eight files will be transferred simultaneously and three of them will be from Small, two from Medium and three from Large chunks.

After transfers start, estimated completion time for each chunk is calculated for every five seconds by dividing the remaining data size to the throughput of the chunk (i.e. the sum of the throughput for all channels for a given chunk). When the transfer of all files in a chunk is completed, the channels of the finished chunk are given to a chunk whose estimated completion time is the largest such that all channels will be used until whole dataset transfer is completed. Following the above example, if Large chunk finishes first, three channels from Large chunk will be given to other chunks based on estimated remaining times such that there will always be eight file transfers until all chunks are finished.

\begin{algorithm}[t]
\scriptsize
\centering
\caption{--- Pro-Active Multi-Chunk (ProMC) Scheduling} \label{alg:pmc}
\begin{algorithmic}[1]
\vspace{1mm}
\Statex
\Function{transfer}{source, destination, BW, RTT, maxCC}
	\State $BDP = BW*RTT$
	\State $allFiles = fetchFilesFromServer()$
	\State $chunks = partitionFiles(allFiles,BDP)$
	\For{$i=0; i<chunks.length; i++$}
		\State $weights[i] =  \delta_{i} * chunks[i].size$  \label{pmc:channelWeights}
		\Comment{Weight of chunk}
		\State $totalWeight = totalWeight+ weights[i]$ 
	\EndFor
	\For{$i=0; i<chunks.length; i++$} \label{pmc:channelAllocation}
		\State $weights[i] = \frac{weights[i]}{totalWeight}$
		\Comment{Proportional weight of chunk}
		\State $chunk[i].concurrency =  \left\lfloor weights[i] * maxCC \right\rfloor $
	\EndFor
	\For{$i=0; i<chunks.length; i++$}
		\State $transfer(chunks[i])$\label{pmc:scheduleall}
		\Comment{Run in the background}
	\EndFor
\EndFunction
\end{algorithmic}
\end{algorithm}

\subsection{Pro-Active Multi-Chunk (ProMC) Scheduling} \label{sec:promc}

The way MC algorithm distributes channels among chunks may lead to sub-optimal allocations because it does not consider weights of the chunks. That is, even if one chunk constitutes 90\% of the entire dataset, it will still be given the same number of concurrency with other chunks under round-robin scheduling. Moreover, round-robin distribution does not consider the fact that not all chunks achieve the same throughput when the same concurrency level is given. Figure~\ref{fig:param-effect-xsede} shows that Small chunk obtains much less throughput than larger chunks at the same concurrency level. Thus, round-robin channel distribution may cause load balancing issues for data channels. Pro-Active Multi-Chunk (ProMC) algorithm distributes the available concurrency level among chunks by considering both weight and types of chunks to address possible load balancing issues.

Channel allocation in the ProMC approach is demonstrated in Algorithm~\ref{alg:pmc}. ProMC considers the type of a chunk and size of chunk when calculating weight for a chunk (line~\ref{pmc:channelWeights} in Algorithm~\ref{alg:pmc}).  $\delta_i$ is coefficient vector that is used to give higher priority to smaller chunks since they tend to yield much less throughput for the same concurrency levels as can be seen in Figure~\ref{fig:xsede_ppq+p+cc} and~\ref{fig:loni_ppq+p+cc}. Although throughput gap between chunks are different at different concurrency levels, it is a general trend that the gap is much higher when small concurrency values are used. For example, 1MB files obtain around 600 Mbps throughput when concurrency level is set to 2 in Figure~\ref{fig:xsede_ppq+p+cc}. For the same concurrency level, 10GB files achieve ~5 Gbps throughput. The difference becomes much less when concurrency level is 4 or 8. Since MC algorithm almost equally distributes concurrency among channels, for the most cases (unless $maxCC > 12)$, each chunk is allotted four or less channels. Hence, we define $\delta$ value for each chunk reversely proportional to their performance under small concurrency level. We use $\delta$ vector \{6,3,2,1\} for \{Small, Medium, Large, Huge\} set in the rest of the experiments. Even though the defined $\delta$ vector may not be optimal, it still allows to assign more concurrency values to Small chunk. Furthermore, we introduce ``online channel re-allocation mechanism" to adjust channel allocations based on real time performances.

Online channel re-allocation helps to correct initial sub-optimal allocations by re-allocating channels from fast chunks to slow chunks. It does this by comparing estimated completion time of each chunk periodically (by default, it is set to check every five seconds). If a chunk's estimated completion time is significantly smaller than another one's for three consecutive periods, then a channel is reassigned from fast chunk to slow chunk. Since different chunks may use different parallelism values, channel re-allocation may need to close the current channel and establish a new one, which causes an overhead. So, the threshold must be chosen carefully when comparing completion time differences. We set the threshold such that slow chunk has to be expected to run at least two times longer than fast chunk. Moreover, rather than deciding on channel allocation after each period, ProMC waits three periods to avoid making incorrect estimations caused by transient events.

\section{Evaluation}
\label{sec:results}
\subsection{How Many Chunks to Create?}

\begin{figure}
\begin{centering}
 \includegraphics[keepaspectratio=true,angle=0,width=60mm]{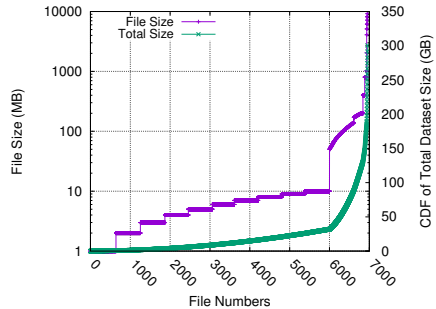}
\caption{File size distribution of dataset used in analysis of creating different number of chunks.} \label{fig:mixed_dataset}
\end{centering}
\end{figure}

\begin{figure*}[!htb]
\begin{center}
\subfigure[Single Chunk (SC)]{
\includegraphics[keepaspectratio=true,angle=0,width=59mm] {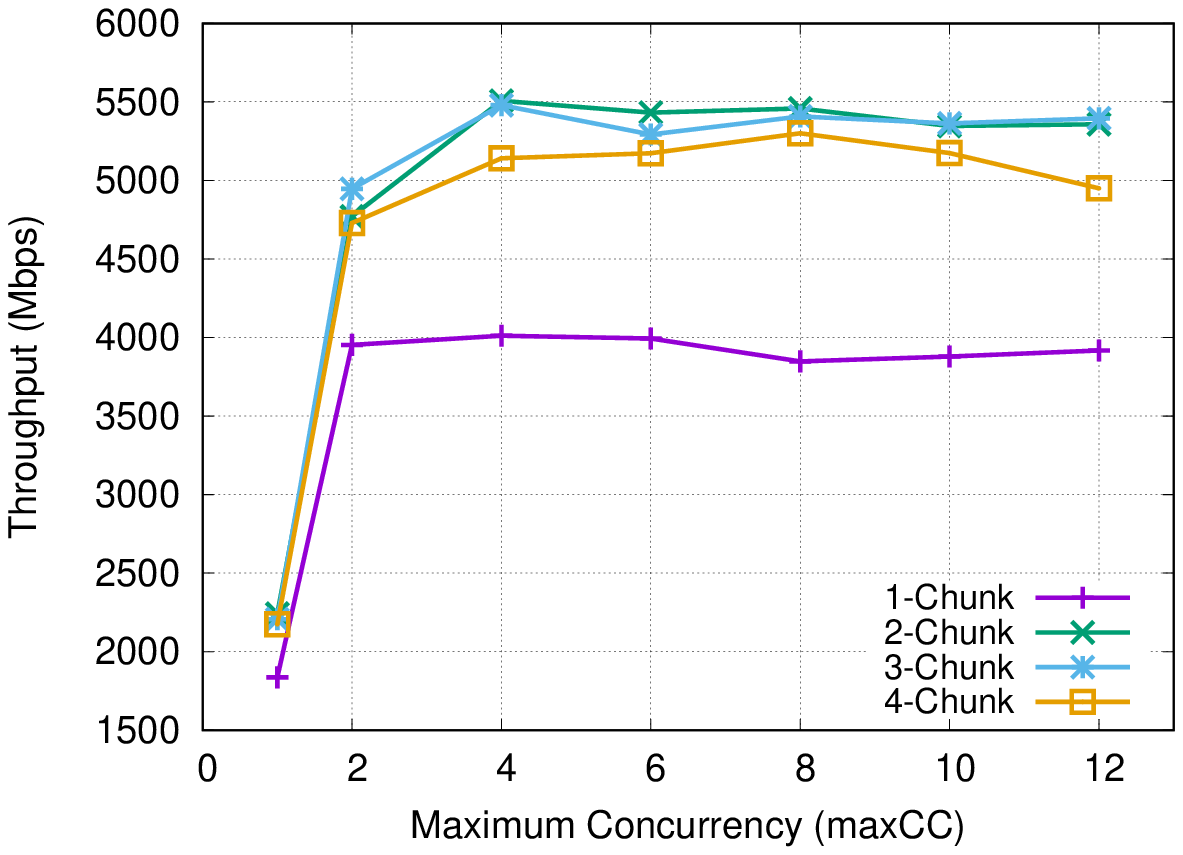}
\label{fig:xsede_reverse}}
\hspace{-4mm}
\subfigure[Multi-Chunk (MC)]{
\includegraphics[keepaspectratio=true,angle=0,width=59mm] {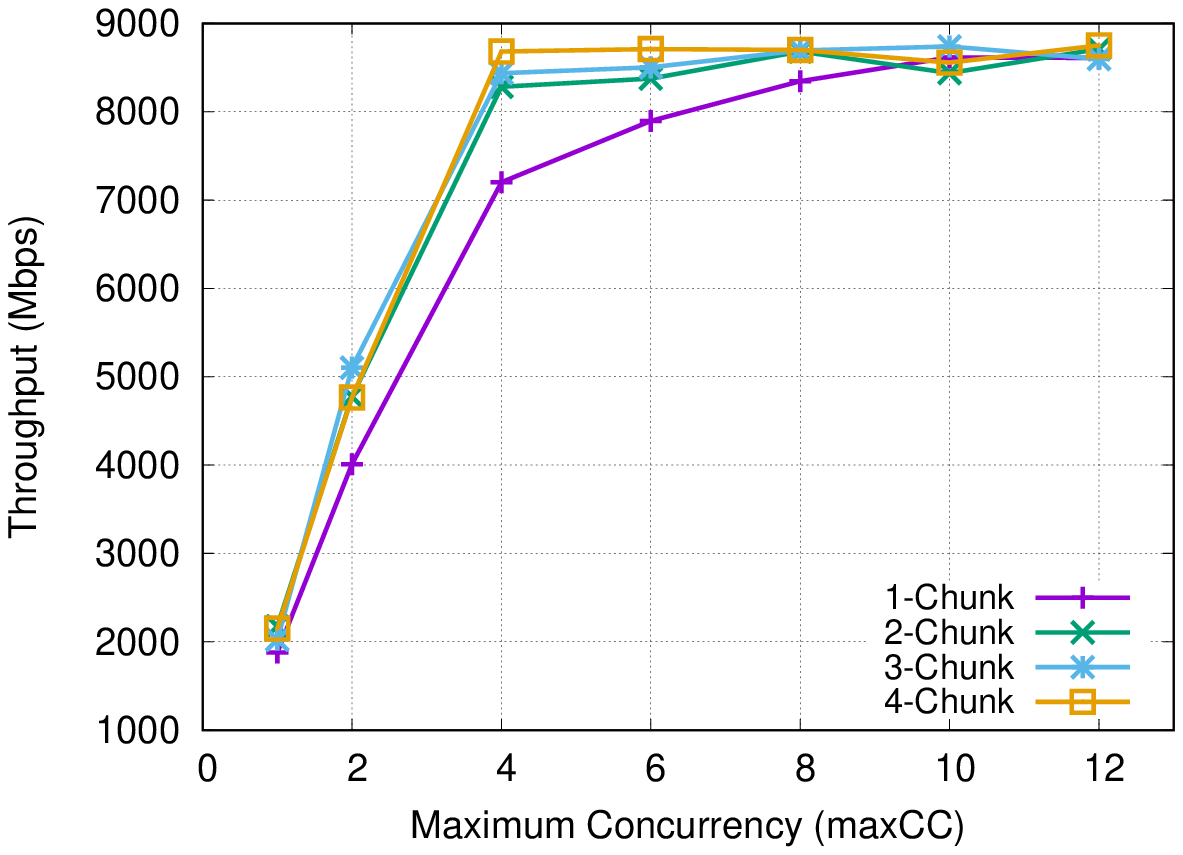}
\label{fig:xsede}}
\hspace{-4mm}
\subfigure[Pro-Active Multi-Chunk (ProMC)]{
\includegraphics[keepaspectratio=true,angle=0,width=59mm] {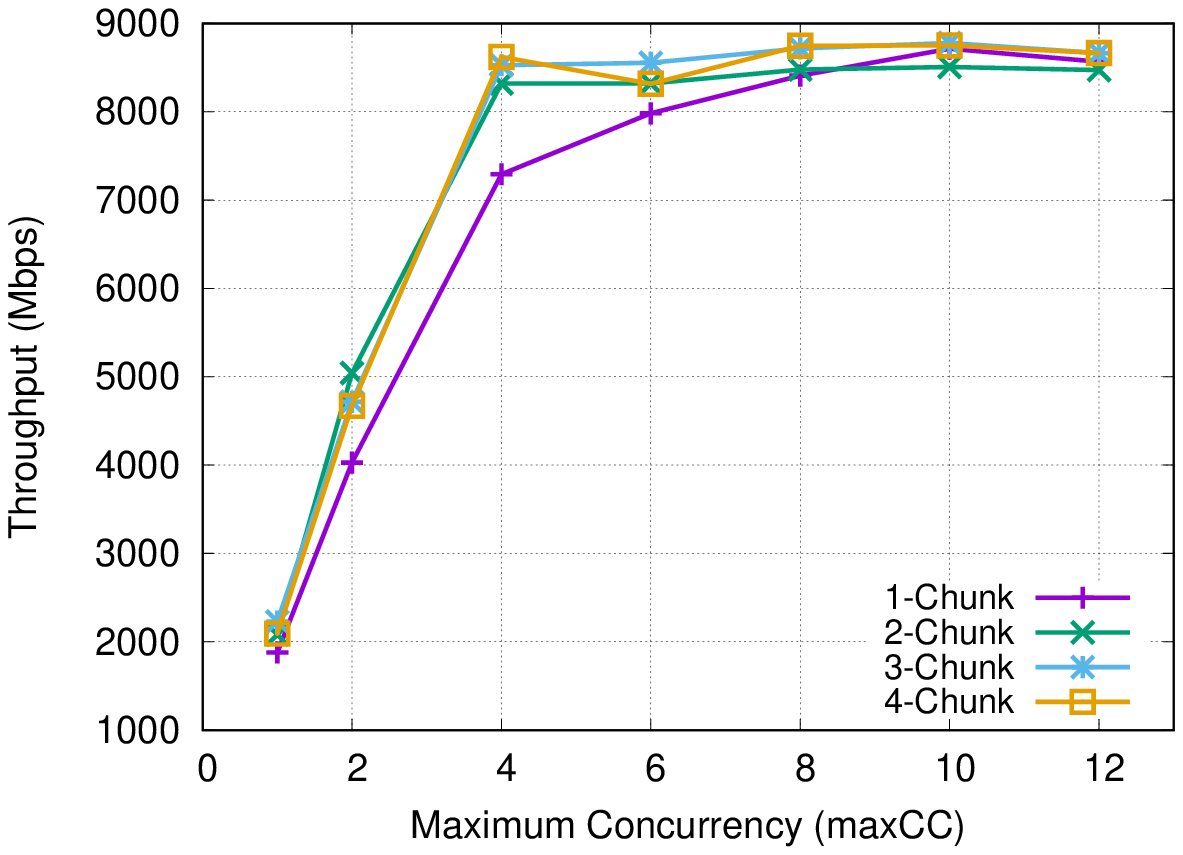} \label{fig:didclab}}
\caption{Impact of creating different number of chunks in Wide Area Network (Stampede-Comet).}
\label{fig:xsede_chunkcount}
\end{center}
\end{figure*}

\begin{figure*}[!htb]
\begin{center}
\subfigure[Single Chunk (SC)]{
\includegraphics[keepaspectratio=true,angle=0,width=59mm] {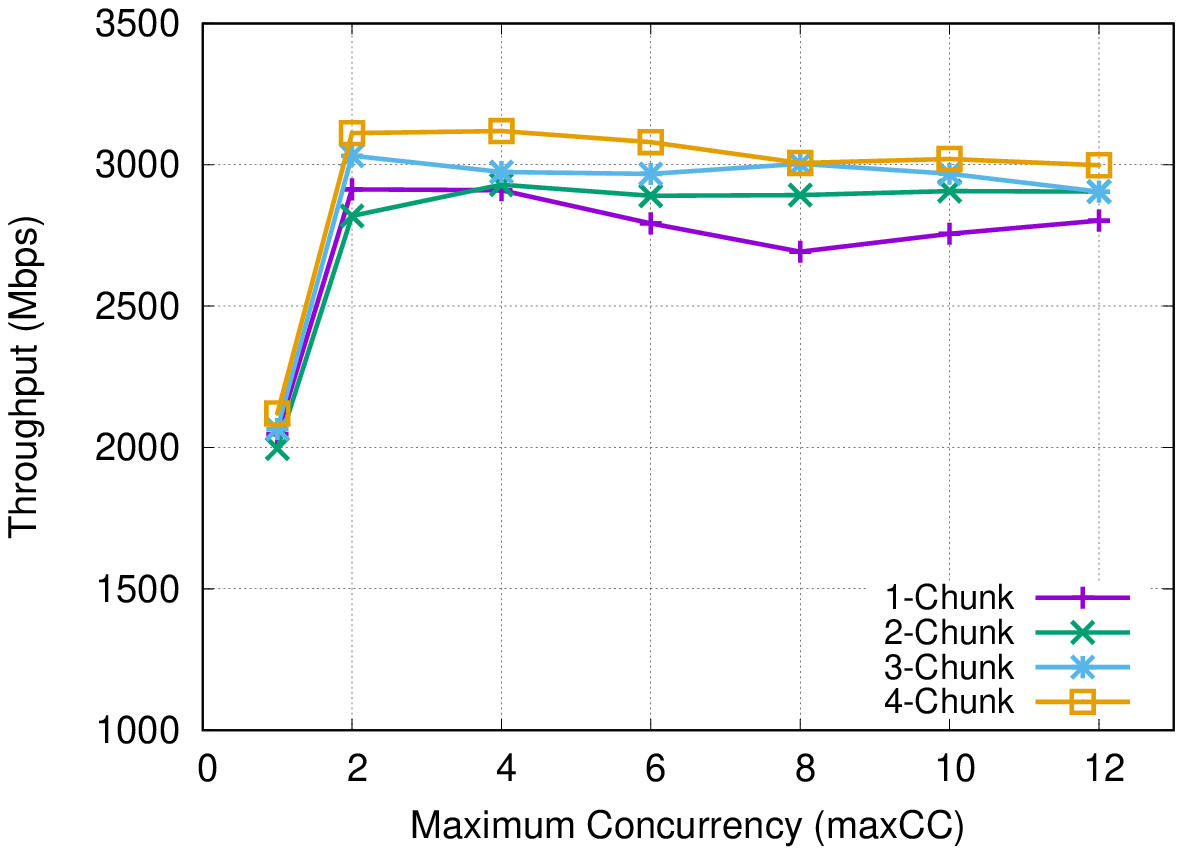}
\label{fig:xsede_reverse}}
\hspace{-4mm}
\subfigure[Multi-Chunk (MC)]{
\includegraphics[keepaspectratio=true,angle=0,width=59mm] {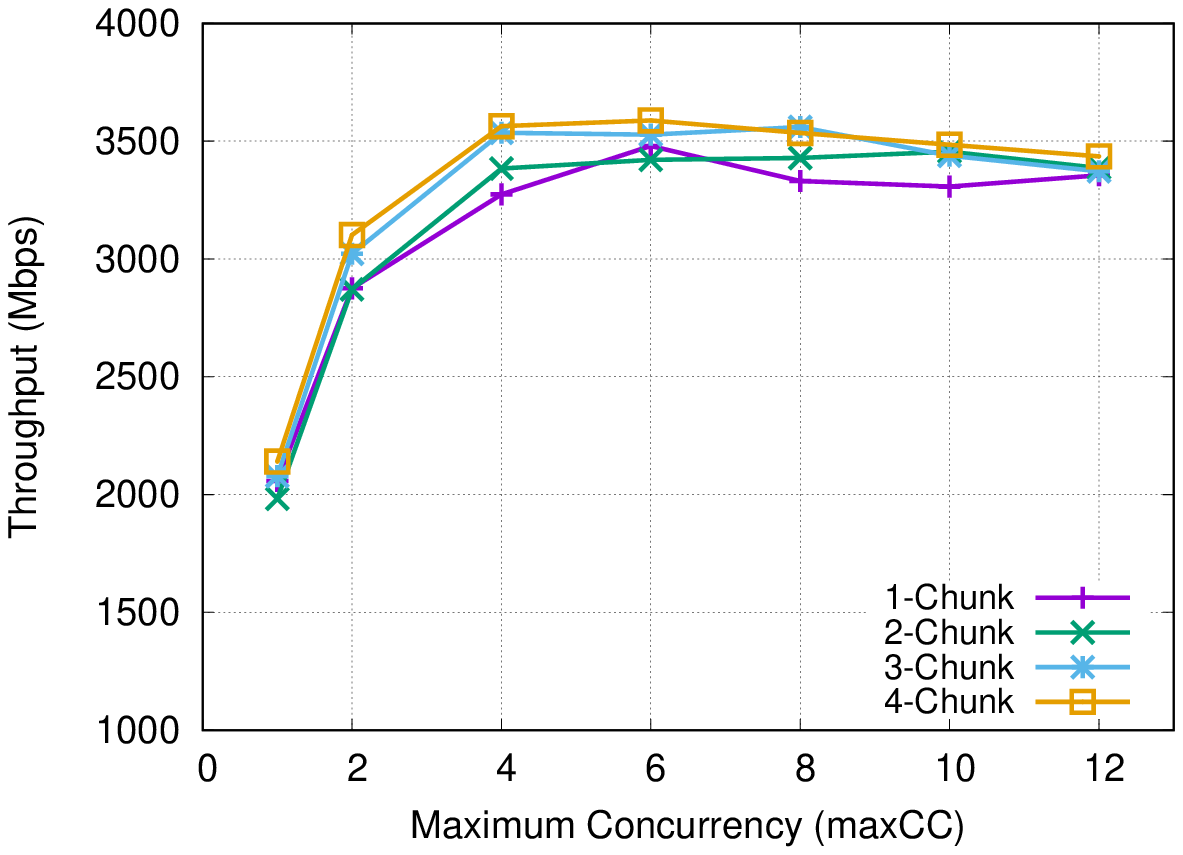}
\label{fig:xsede}}
\hspace{-4mm}
\subfigure[Pro-Active Multi-Chunk (ProMC)]{
\includegraphics[keepaspectratio=true,angle=0,width=59mm] {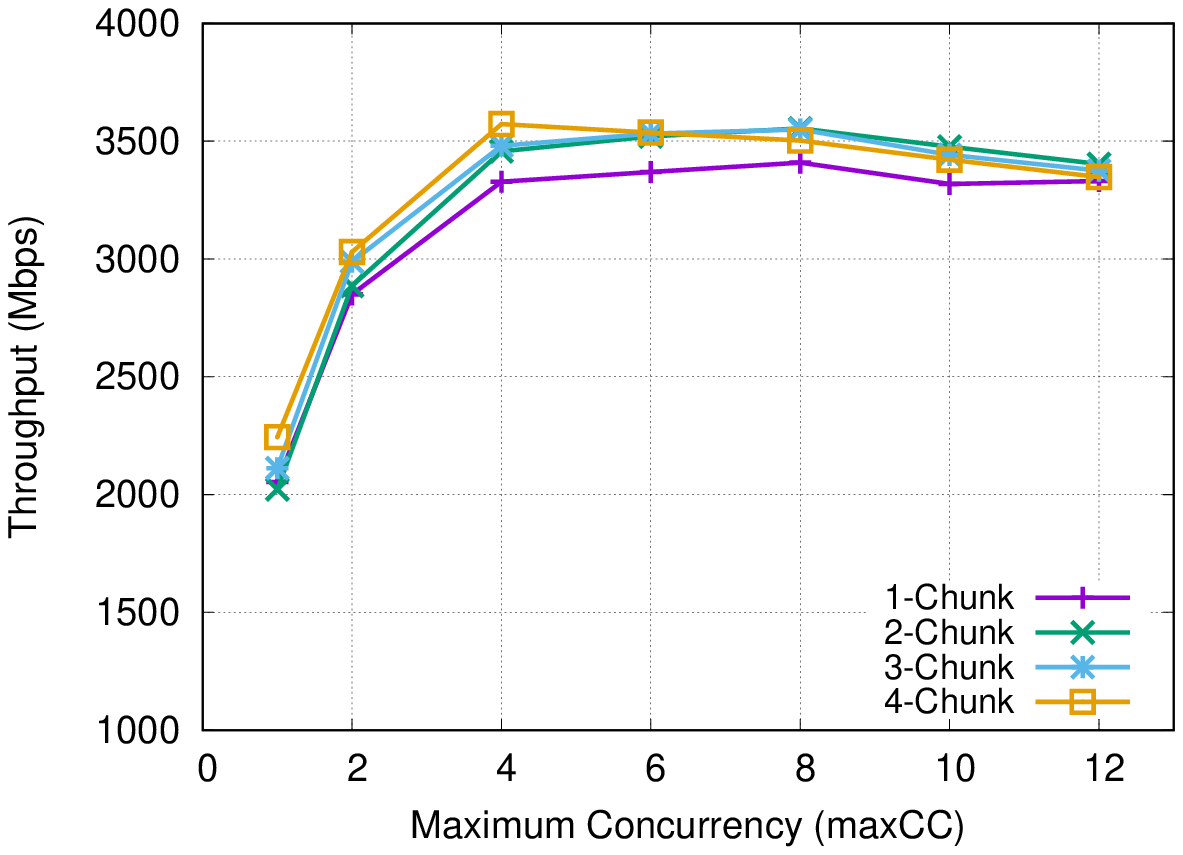} \label{fig:didclab}}
\caption{Impact of creating different number of chunks in local-area network.}
\label{fig:lan_chunkcount}
\end{center}
\end{figure*}

In order to measure the effect of creating different number of chunks from a mixed dataset, we have compared the performance of SC, MC, and ProMC under different chunk counts. Figure~\ref{fig:xsede_chunkcount} and~\ref{fig:lan_chunkcount} shows the results of creating different number of chunks using the same dataset in wide-area (Stampede-Comet) and local-area networks. We have used a dataset with mixed file sizes as shown in Figure~\ref{fig:mixed_dataset}. File sizes range from 1 MB to 9.2 GB and total dataset size is 300.5 GB. 

In the WAN experiment (Figure~\ref{fig:xsede_chunkcount}), the throughput achieved by SC increases up to concurrency level 4, and then becomes steady for all chunk counts. This is because concurrency equation in Algorithm~\ref{alg:parameterEstimation} returns small concurrency value (it is 2 in this experiment) for Medium, Large, and Huge chunk types even if $maxsCC$ is large. The reason behind small concurrency value is that concurrency equation in Algorithm~\ref{alg:parameterEstimation} is limited by $maxCC$ until $maxCC \leq 2$ then it is limited by $Max(\frac{BDP}{avgFileSize}, 2)$.  Equation~\ref{eq:two} shows that $\frac{BDP}{avgFileSize}$ is always less than $20*RTT$ for Medium chunk type and it will be even smaller for Large and Huge chunks as average file size grows. $20*RTT$ will be smaller than 2 when $RTT < 100 ms$ which is the case for our wide are network experiment. Thus, most chunks self-limits their use of concurrency as our heuristic calculations estimate that use of small concurrency will suffice to achieve the close to maximum throughputs. This assumption may not hold true if other conditions who are not considered in our calculations play a major role in determining transfer throughput such as disk throughput and background traffic.

\begin{equation*}
 x=avgFileSize,~~~y = \frac{BDP}{avgFileSize} = \frac{BW*RTT}{x} \\
\label{eq:one}
\end{equation*}

For Medium chunk $\frac{BW}{20} < avgFilesize = x \leq \frac{BW}{5} $, so:

\begin{equation}
\begin{split}
&\frac{BW}{20} < x \leq \frac{BW}{5} \textnormal{, Replacing x with } \frac{BW*RTT}{y}\\
&=\frac{BW}{20} < \frac{BW*RTT}{y} \leq \frac{BW}{5}\\
&=\frac{1}{20} < \frac{RTT}{y} \leq \frac{1}{5}\\
&=5 \leq  \frac{y}{RTT} < 20\\
&=5*RTT \leq y < 20*RTT
\label{eq:two}
\end{split}
\end{equation}

Since Small chunk type can benefit from increasing level of $maxCC$, partition techniques with Small chunk type (i.e., 2, 3, and 4-chunk) achieve up to 20\% more throughput than 1-chunk partitioning which treats whole dataset as one chunk so does not have separate Small chunk. This is because concurrency calculation depends on average file size and taking whole dataset as single chunk makes the chunk's average file size larger than Small-only chunk. On the other hand, 2, 3, and 4-chunk partitions achieve similar performance as they all benefit from having Small chunk and all other chunk types (Medium, Large, and Huge) uses similar parameter values so obtains very close transfer throughputs. Although Small chunk uses as many as $maxCC$ value, increasing $maxCC$ does not affect increase overall throughput for 2, 3, or 4-chunk partitions considerably after $maxCC=4$ since throughput gain starts decreasing and its impact over whole dataset becomes negligible. 

As opposed to SC, MC and ProMC are able to achieve around 9 Gbps throughput by (i) using maximum allowed concurrency level, $maxCC$, and (ii) transferring multiple chunks simultaneously in order to minimize the effect of small files over whole dataset. All partitioning approaches perform similar for MC and ProMC when $maxCC \geq 8$. However, when $maxCC$ is smaller, 1-chunk partition achieves up to 20\% less than other partitions. This is because when concurrency level is large enough, the impacts of parallelism and pipelining diminish as discussed in~\cite{tcc16}. 
It is also worth to note that 2, 3, and 4-chunk partitions in MC and ProMC algorithms perform similarly and they all can achieve maximum throughput with as little as four channels. Even though Small chunk requires much larger concurrency level to optimize the transfer throughput, MC and ProMC can achieve the highest throughput using smaller number of channels.

The results from local area network experiments are shown in Figure~\ref{fig:lan_chunkcount}. In the LAN experiments, we used half of dataset compared to the WAN experiments, since the achievable throughput is smaller. However, we kept the file distributions the same. Using one set of parameters for whole dataset (aka 1-chunk) obtains slightly worse throughput than 2, 3, and 4-chunk partitions for SC algorithm similar to WAN experiments. However, since there is not much room to optimize, the difference becomes less than 10\%. Similarly, 1-chunk obtains slightly less throughput than other partitioning cases for MC and ProMC. 2, 3, and 4-chunk partitioning performs very close to each other and their throughput decreases a bit when the maximum concurrency is set to larger than 4 which can be explained due to disk contention as the storage is backed by only five servers.

\begin{figure}
\begin{centering}
 \includegraphics[keepaspectratio=true,angle=0,width=60mm]{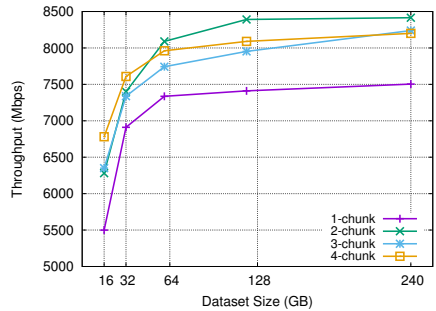}
\caption{The effect of dataset size over the performance of different partitioning techniques (MC with maxcc=6).} \label{fig:dataset-size}
\end{centering}
\end{figure}

We also analyzed how partitioning techniques affect the transfer throughput for different dataset sizes. Figure~\ref{fig:dataset-size} shows the results when five different dataset sizes are used for MC with fixed maximum concurrency value of 6. The dataset used in this experiment contains files for all file types (Small, Medium, Large, Huge) with close-to equal sizes. As an example, when dataset size is 16 GB, all four chunk types has around 4 GB size. When maximum chunk size is given less than four, then the created chunks will not have same size. 1-chunk partitioning again performs worse than others due to incapability of fine tuning parameter values. While 4-chunk partitioning performs well for small datasets, it is outperformed by 2 and 3-chunk partitions as the dataset size increases. This is mainly because 2-chunk partitioning allocates more channels to Small chunk for the same concurrency level as MC evenly distributes channels among chunks. For example, when 2-chunk partitioning is used to transfer 16 GB of data, Small chunk will have 4 GB size and the rest of the dataset will be combined into a single chunk and will be 12 GB in size. Since MC's round-robin distribution is oblivious to chunk sizes, it allocates the same number of channels to each chunk, which leads to more fair share channel distribution since Small chunk requires more channels to achieve the same transfer throughput as larger chunks. Additionally, greater number of chunks means more channel reallocation from finished chunks to slow chunks which may cause some delay if parallelism values of chunks are different as discussed in Section~\ref{sec:sc}.

As a result, transferring whole dataset as one chunk is not optimal for both WAN and LAN transfers except when MC or ProMC are used with large concurrency values. Moreover, creating more than two chunks helps only when the dataset is too small. While one can easily adjust the number of chunks based on the dataset size, we used 2-chunk partitioning in the rest of the experiments as we try to optimize large data transfers in this work. 

\begin{figure*}[!htb]
\begin{center}
\subfigure[Dark Energy Survey]{
\includegraphics[keepaspectratio=true,angle=0,width=59mm] {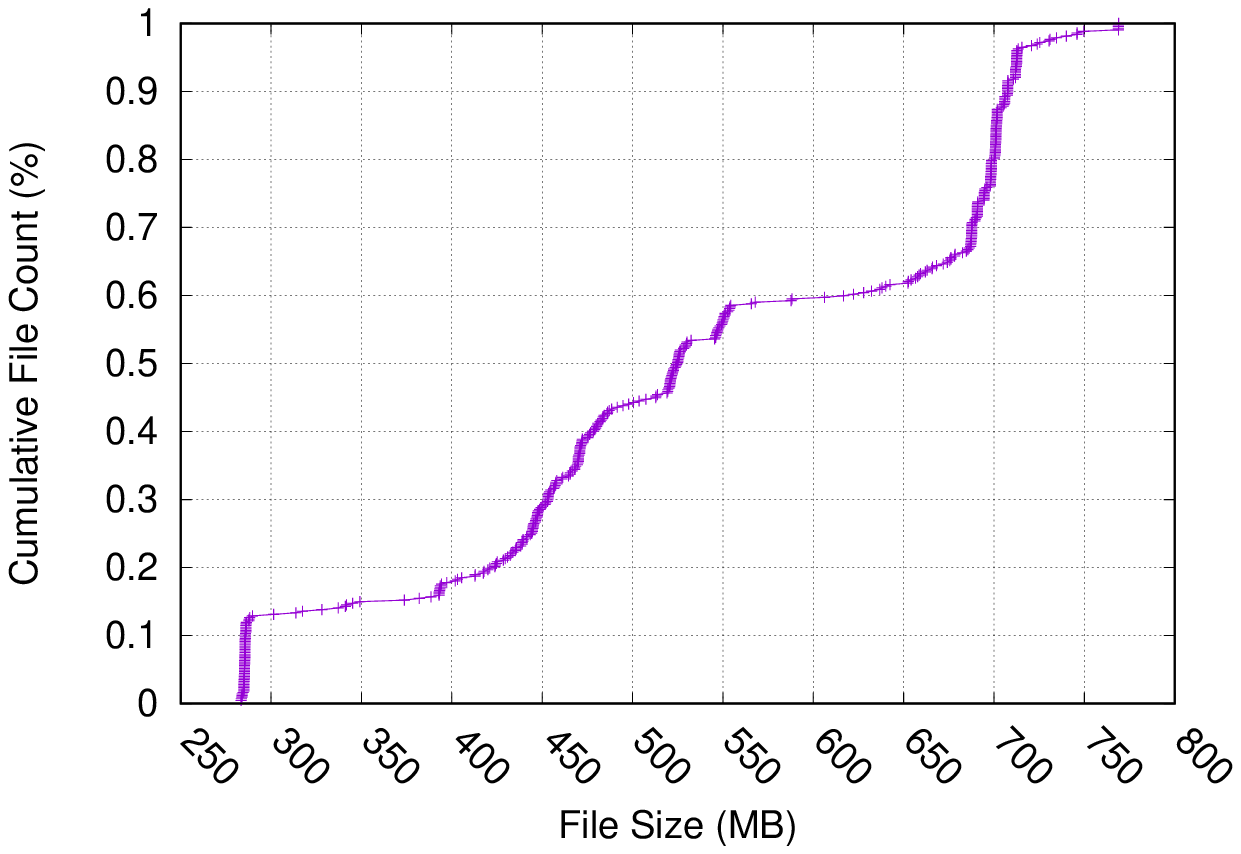}
\label{fig:file_dist_dark}}
\hspace{-4mm}
\subfigure[Genome Sequencing]{
\includegraphics[keepaspectratio=true,angle=0,width=59mm] {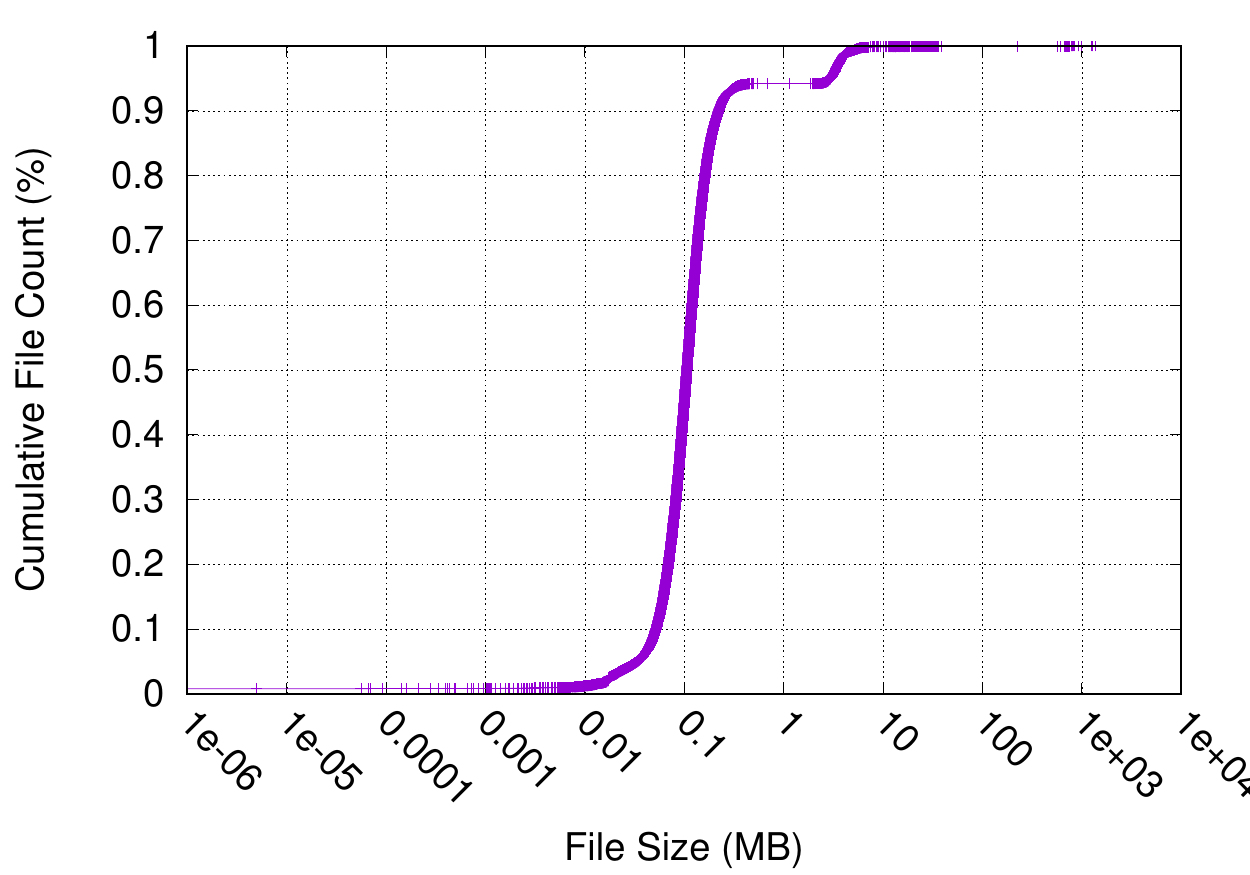}
\label{fig:file_dist_bio}}
\hspace{-4mm}
\subfigure[Mixed]{
\includegraphics[keepaspectratio=true,angle=0,width=59mm] {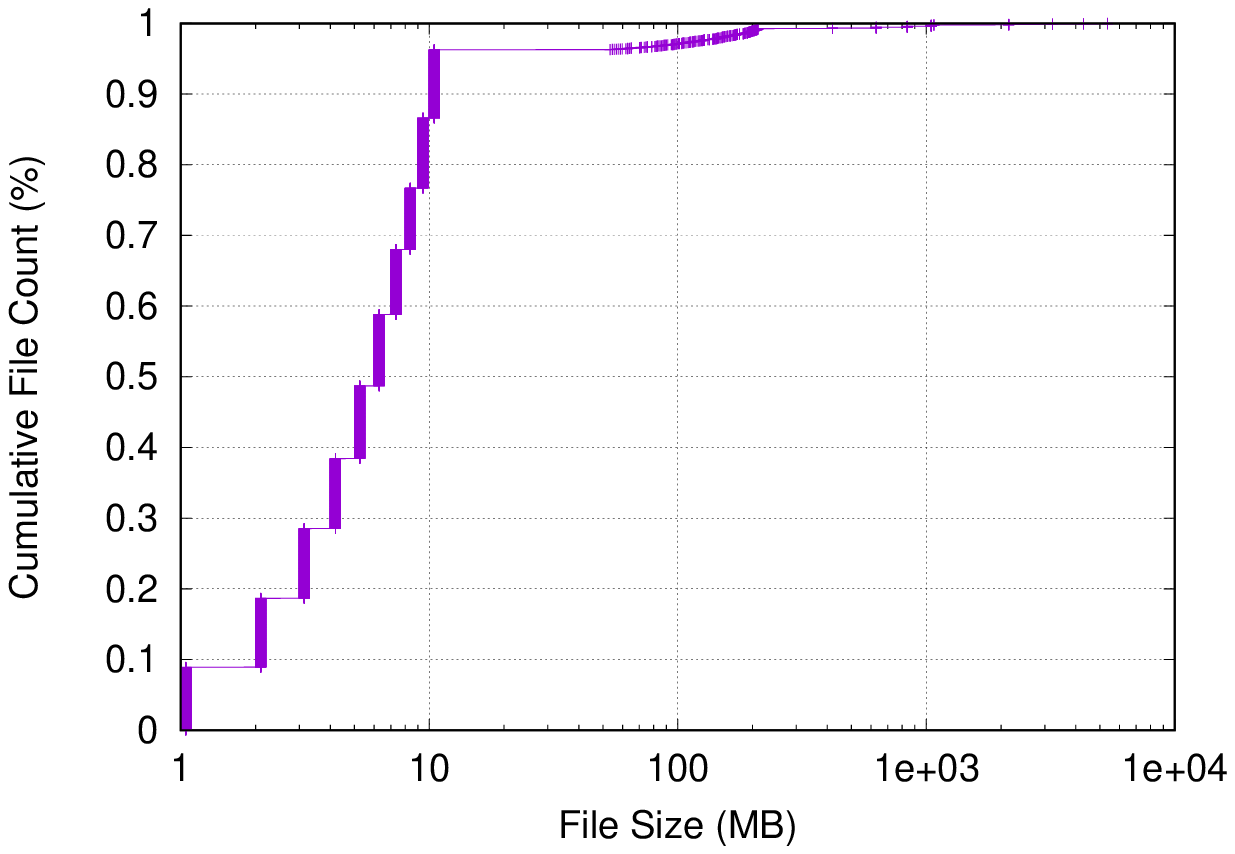} \label{fig:file_dist_mixed}}
\caption{File size distribution of the datasets used in performance evaluation of algorithms.}
\label{fig:file_dist_all}
\end{center}
\end{figure*}

\subsection{Performance Comparison}

\begin{table}
\begin{centering}
\resizebox{\textwidth}{!}{
\begin{tabular}{ |c| c| c| c| c|}
\hline
 \multirow{2}{*}{\bf Specs} &\multicolumn{3} {|c|} {XSEDE} & \multirow{2}{*} {LAN}\\
& BlueWaters-Stampede & Stampede-Comet & SuperMIC-Bridges& \\
\hline
{\bf Bandwidth (Gbps)} &  3x10 & 10  & 10 & 10 \\
\hline
{\bf RTT (ms)} & 32 & 40 & 45 & 0.2  \\
\hline
{\bf TCP Buffer Size (MB)} & 32 & 32 & 4 & 1  \\
\hline
{\bf BDP (MB)} & 40 & 50 & 56 & 0.25 \\
\hline
{\bf File System} & Lustre & Lustre & Lustre  & GlusterFS\\
\hline
\end{tabular}}
\caption{Network specifications of the test environment.} \label{tab:system-spec2}
\end{centering}
\end{table}

\begin{figure*}[!htb]
\begin{center}
\subfigure[BlueWaters-Stampede]{
\includegraphics[keepaspectratio=true,angle=0,width=59mm] {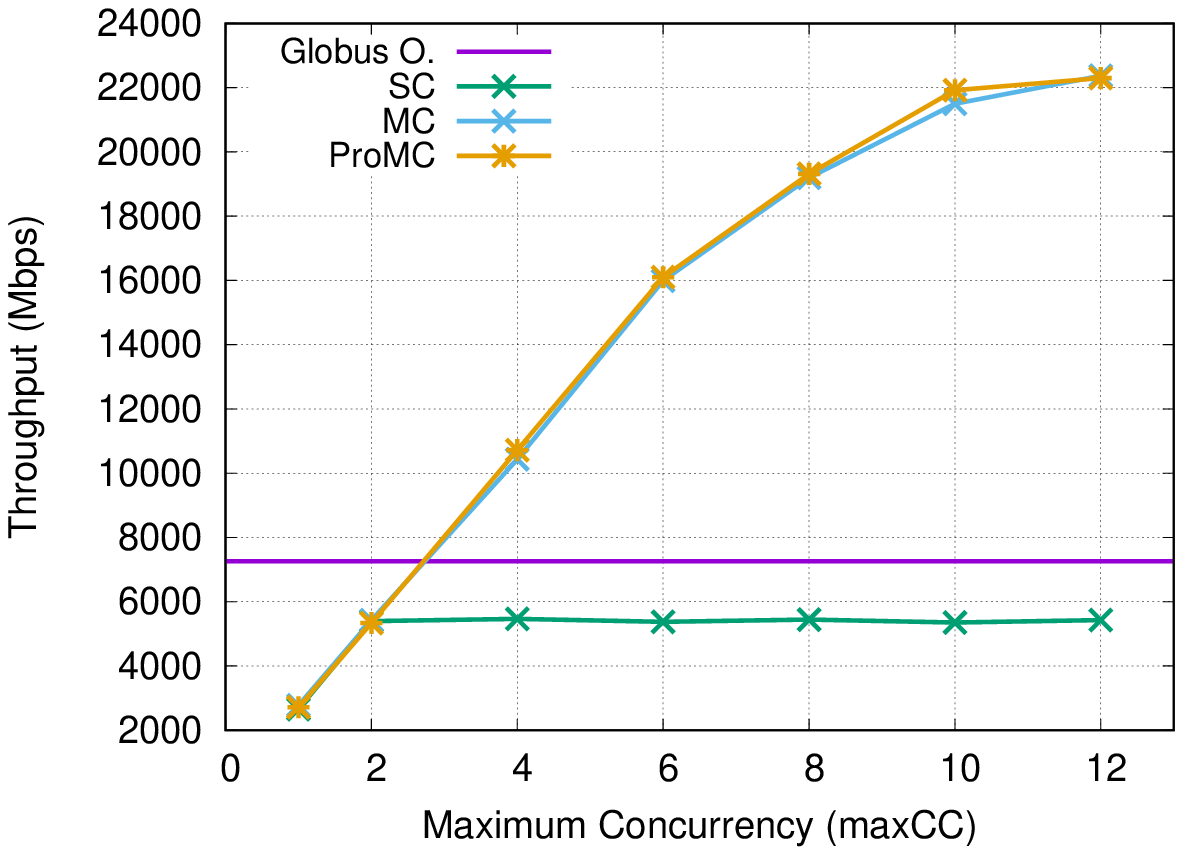}
\label{fig:dark-energy-bw-stampede}}
\hspace{-4mm}
\subfigure[Stampede-Comet]{
\includegraphics[keepaspectratio=true,angle=0,width=59mm] {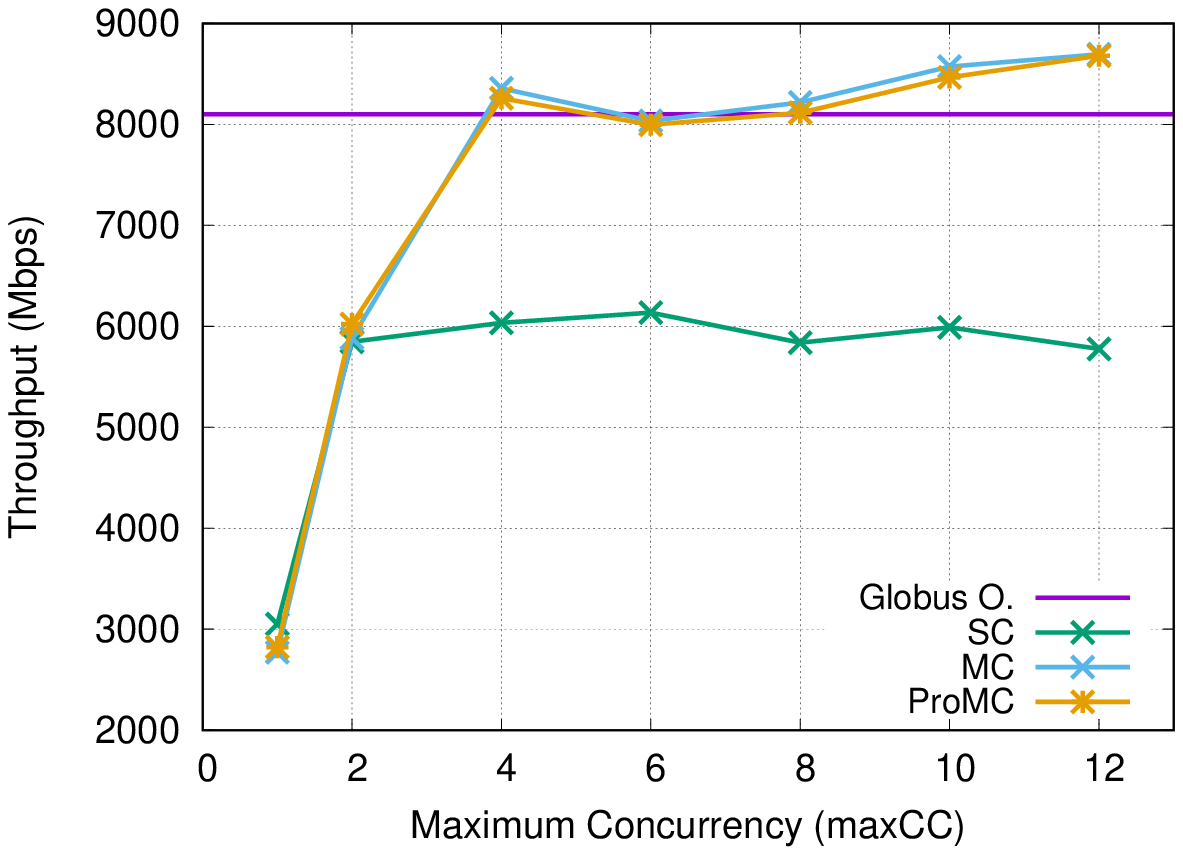}
\label{fig:dark-energy-stampede-comet}}
\hspace{-4mm}
\subfigure[SuperMIC-Bridges]{
\includegraphics[keepaspectratio=true,angle=0,width=59mm] {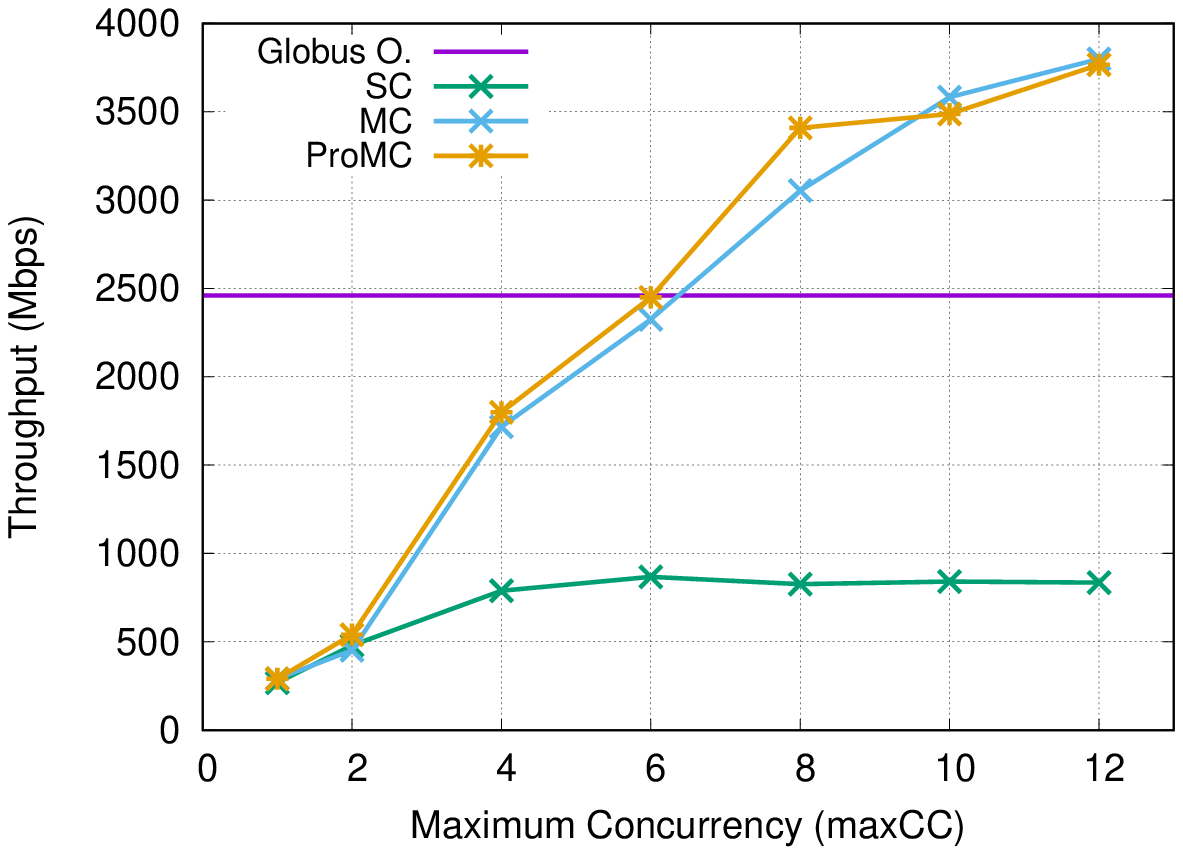} \label{fig:dark-energy-supermic-bridges}}
\caption{Performance comparison of algorithms with Dark Energy Survey dataset.}
\label{fig:dark-energy-all}
\end{center}
\end{figure*}

\begin{figure*}[!htb]
\begin{center}
\subfigure[BlueWaters-Stampede]{
\includegraphics[keepaspectratio=true,angle=0,width=59mm] {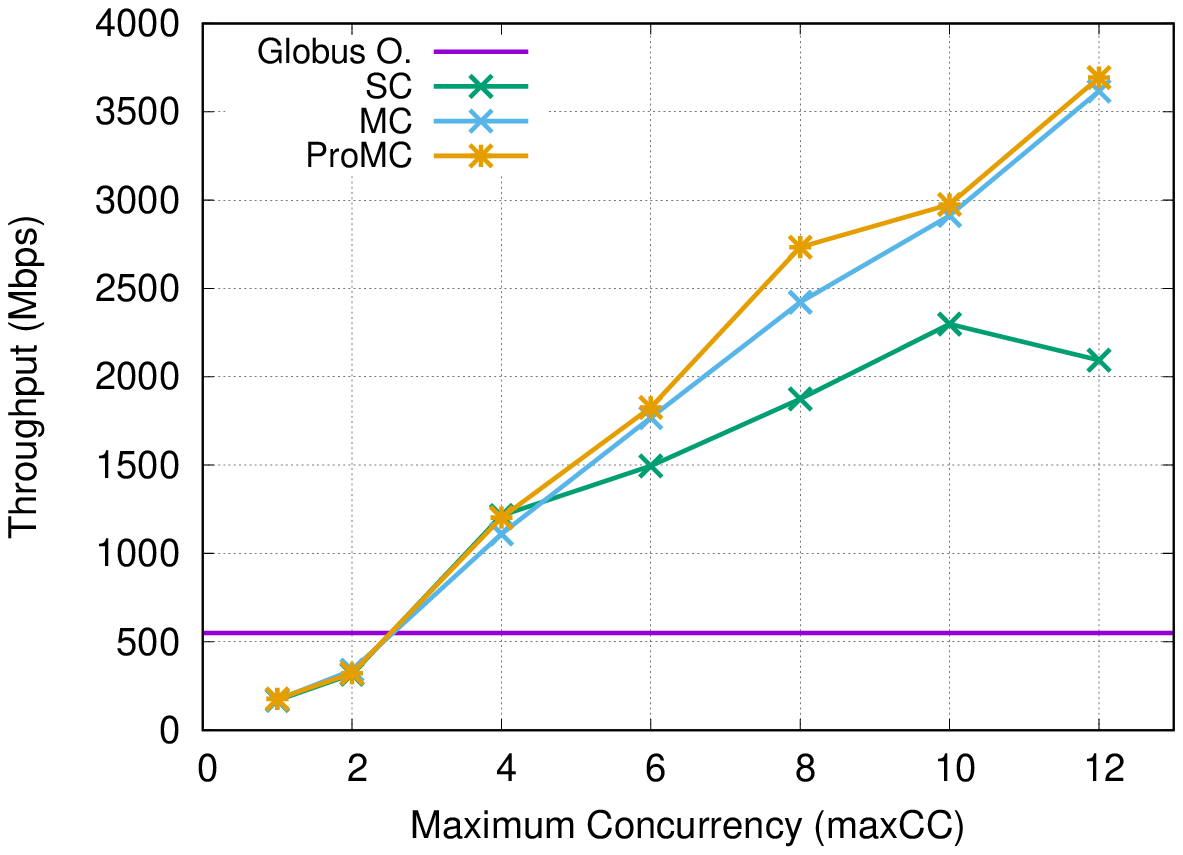}
\label{fig:bio-bw-stampede}}
\hspace{-4mm}
\subfigure[Stampede-Comet]{
\includegraphics[keepaspectratio=true,angle=0,width=59mm] {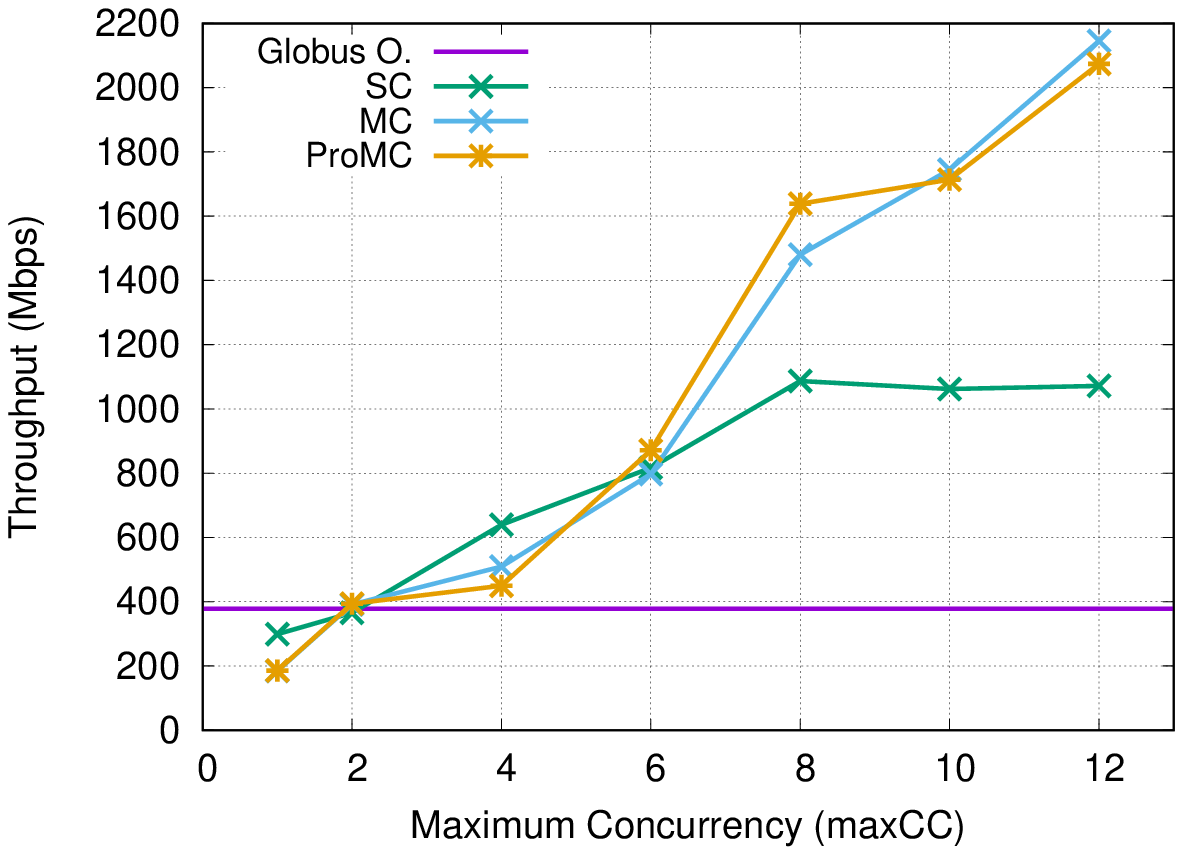}
\label{fig:bio-stampede-comet}}
\hspace{-4mm}
\subfigure[SuperMIC-Bridges]{
\includegraphics[keepaspectratio=true,angle=0,width=59mm] {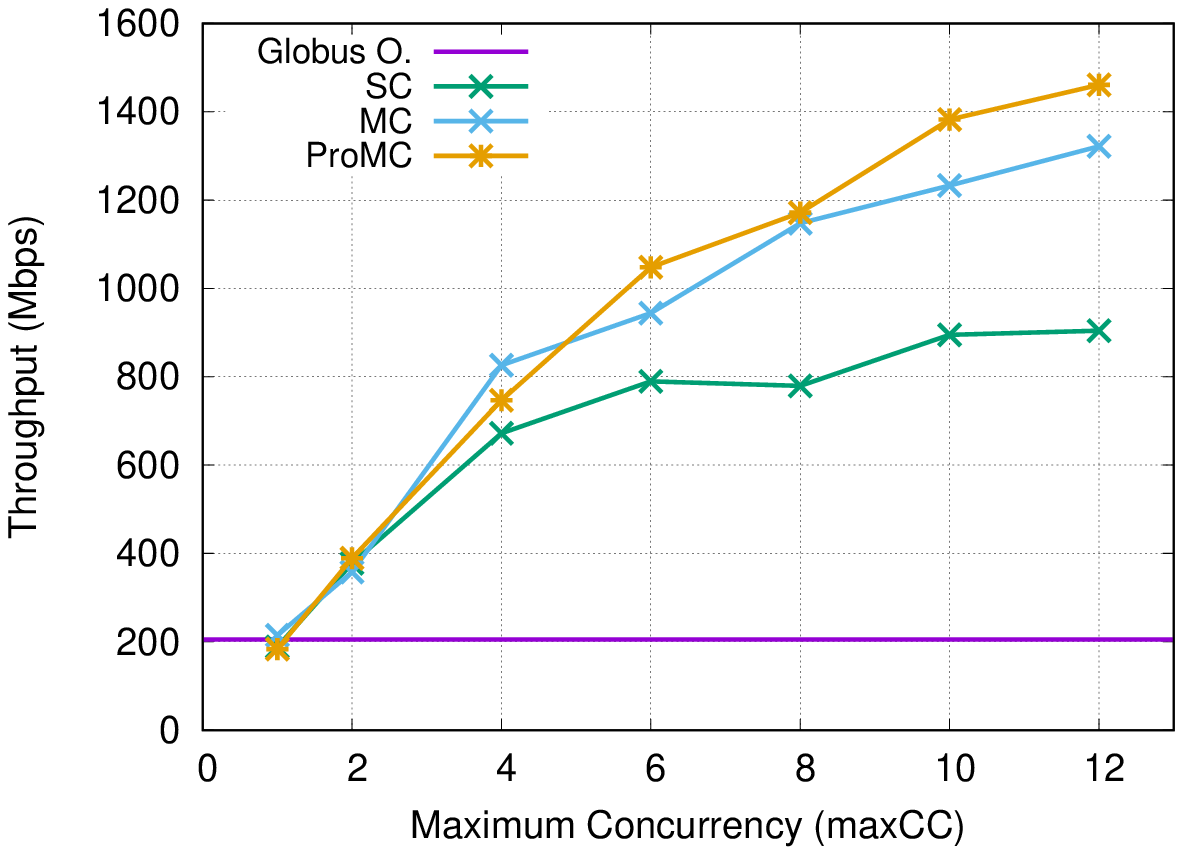}\label{fig:bio-supermic-bridges}}
\caption{Performance comparison of algorithms with Genome sequencing dataset.}\label{fig:bio-all}
\end{center}
\end{figure*}

We tested our experiments on XSEDE~\cite{xsede} wide-area and DIDCLAB local-area networks.
XSEDENet is the high speed network between XSEDE service providers sites around US. Each site is connected with 10/30 Gbps network bandwidth and uses Internet-2 backbone to provide dedicated network capacity. I/O accesses of XSEDE sites are backed by parallel file systems, mostly Lustre. In local area network experiments, we used two VM instances that are backed by GlusterFS.

In XSEDENet, we tested our dynamic protocol tuning algorithms between three site pairs -- BlueWaters-Stampede, Stampede-Comet, and SuperMIC-Bridges -- as specifications given in Table \ref{tab:system-spec2}.
We used three different datasets; Dark Energy Survey~\cite{dark-energy-survey}, genome sequencing~\cite{pacific-bio}, and mixed datasets whose file distributions are shown in Figure~\ref{fig:file_dist_all}. Dark Energy Survey dataset contains files that are collected at the observatory over one day period and consists of 427 files. File sizes are between 250 MB to 750 MB and total size is 212 GB. Genome sequencing dataset is generated by running Falcon~\cite{falcon}, genome assembly toolkit, on a public genome sequencing reads.
Finally mixed dataset is synthetically generated to involve files from all file types (Small, Medium, Large and Huge). There are 6,232 files in the dataset and the file sizes range between 1 MB to 5 GB as shown in Figure~\ref{fig:file_dist_mixed}.

We compared the performance of our three dynamic protocol tuning algorithms with Globus Online~\cite{globusonline} which is a widely adopted data transfer service. 
%
Since Dark Energy Survey dataset only consists of large files and disk I/O throughput of large files tend to be larger than smaller ones, the throughput values we obtained are the largest among all tests. We are able to achieve around 22 Gbps throughput when Multi-Chunk (MC) and Pro-Active Multi-Chunk (ProMC) are used in BlueWaters-Stampede tests as shown in Figure~\ref{fig:dark-energy-bw-stampede}. Single-Chunk (SC) performs the worst since it limits the use of concurrency to smaller values which causes low I/O throughput. The performance of Globus Online differs based on the parameters it selects but we have observed that it selects concurrency and parallelism values to be less than or equal to 4 and 6, respectively. Thus, its average performance stays less than 8 Gbps while the maximum reaches to 8.5 Gbps. It is important to note that the throughputs of MC and ProMC start decreasing after concurrency value 8 because of overloading disk I/O after reaching the capacity. 

\begin{figure*}[!htb]
\begin{center}
\subfigure[BlueWaters-Stampede]{
\includegraphics[keepaspectratio=true,angle=0,width=59mm] {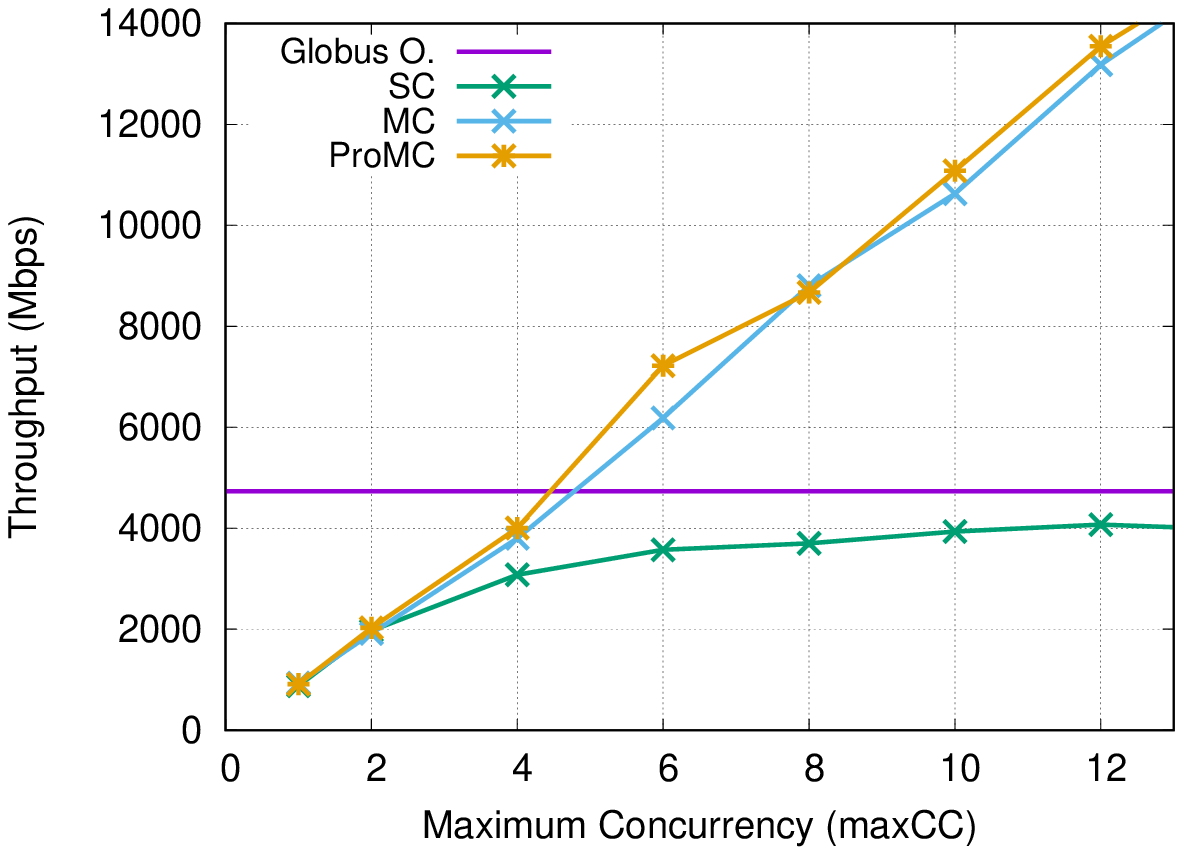}
\label{fig:mixed-bw-stampede}}
\hspace{-4mm}
\subfigure[Stampede-Comet]{
\includegraphics[keepaspectratio=true,angle=0,width=59mm] {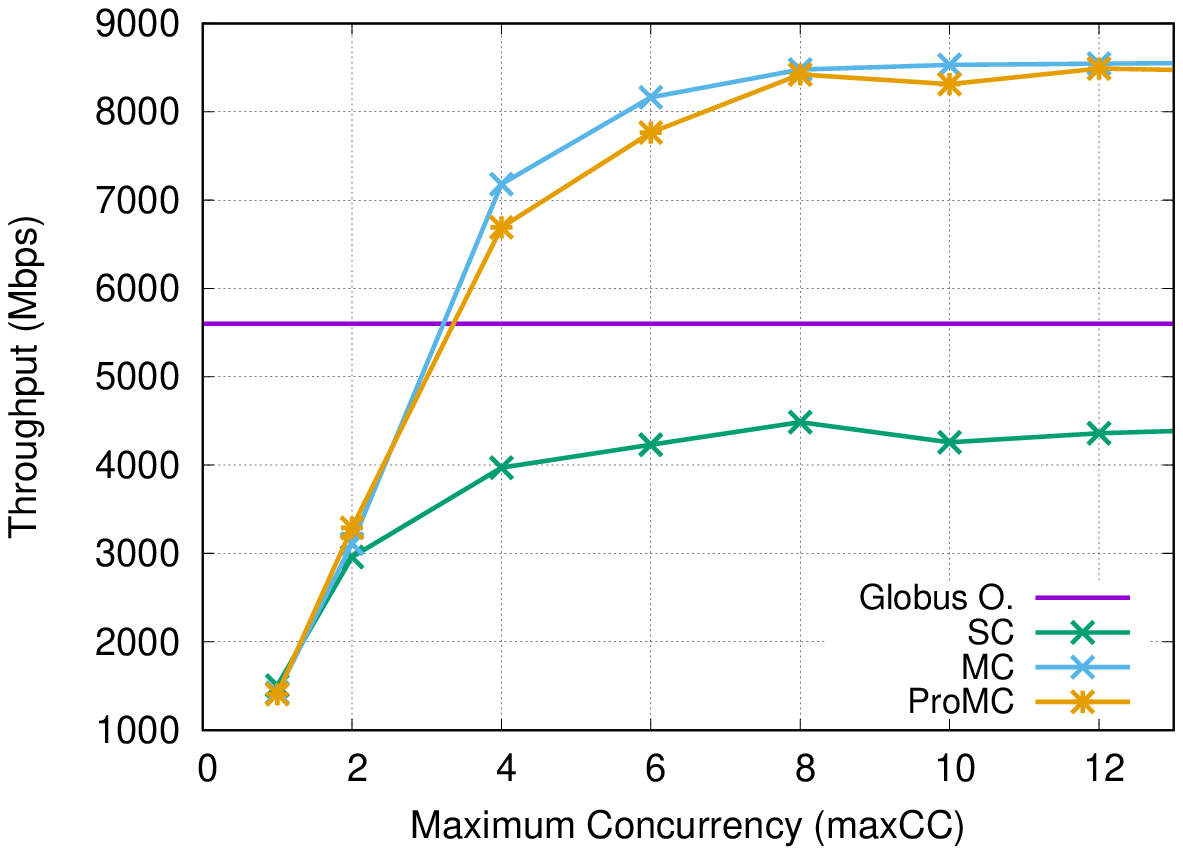}
\label{fig:mixed-stampede-comet}}
\hspace{-4mm}
\subfigure[SuperMIC-Bridges]{
\includegraphics[keepaspectratio=true,angle=0,width=59mm] {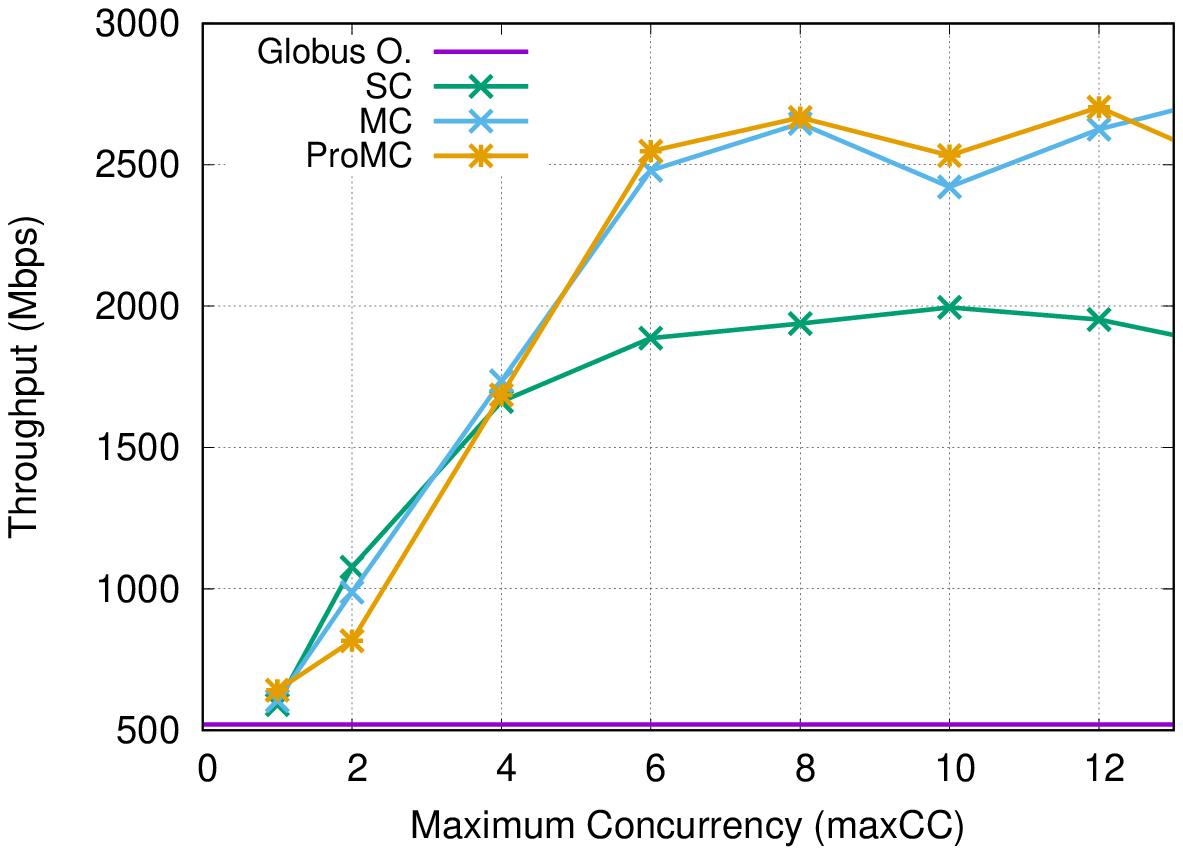} \label{fig:mixed-supermic-bridges}}
\caption{Performance comparison of algorithms with the mixed dataset.}
\label{fig:mixed-all}
\end{center}
\end{figure*}

For Stampede-Comet transfers, Globus Online selects larger values for concurrency and parallelism, and achieves close to 8 Gbps. MC and ProMC, on the other hand, can achieve up to 8.6 Gbps when higher concurrency values are allowed. While the impact of concurrency diminishes as its value increases for BlueWaters-Stampede and Stampede-Comet transfers, it leads to consistent increase in transfer throughput for SuperMIC-Bridges transfers as shown in~\ref{fig:dark-energy-supermic-bridges}. This is due to sub-optimal TCP buffer size settings at SuperMIC as given in Table~\ref{tab:system-spec2}. While it requires over 50 MB of TCP buffer to reach the maximum transfer throughput, current settings only allow 4 MB buffer to be allocated for a transfer. Thus, as we increase the number of concurrency it helps to alleviate buffer size limitations by running multiple file transfers simultaneously and increasing cumulative allocated buffer size for this transfer task. Both MC and ProMC achieve close to 4 Gbps when {\it maxCC} is set to larger values. 

File distribution of genome sequencing dataset is shown in Figure~\ref{fig:file_dist_bio}. There are around 120K files in the dataset and 45\% of files are less than 100 KB and  93\% of files are smaller than 1 MB. Hence the dataset is dominated by very small files although there are several large files up to 13 GB in size. Since average file size of whole dataset is around 500 KB, the throughput values are a lot smaller compared to Dark Energy Survey transfers. In terms of the achieved transfer throughput, MC and ProMC perform similar and obtain 1.5-3.5 Gbps at different site pairs as shown in Figure~\ref{fig:bio-all}. As opposed to Dark Energy Survey dataset transfers, SC performs closer to MC and ProMC since concurrency calculations returns higher values as average file size of dataset is small. 

\begin{figure}
\begin{centering}
 \includegraphics[keepaspectratio=true,angle=0,width=60mm]{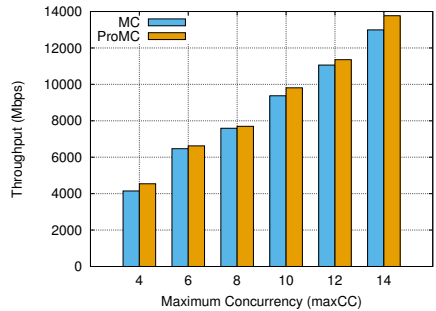}
\caption{Performance comparison of MC and ProMC with the small-file dominated mixed dataset.} \label{fig:mixed_small}
\end{centering}
\end{figure}

Moreover, we compared algorithms using mixed dataset as shown in Figure~\ref{fig:mixed-all}. Mixed dataset contains 6,232 files and file sizes range between 1 MB and 5 GB. Similar to genome sequencing dataset, MC and ProMC perform significantly better than Globus Online. In order to show the difference between MC and ProMC, we doubled the size of small files in mixed dataset and compared them in Figure~\ref{fig:mixed_small}. As the small files dominate the dataset, channel allocation policy becomes more important in order to minimize the effect of Small chunk over average transfer throughput. Otherwise, large chunks finish quickly and small chunks will decrease the overall throughput. Although MC also tries to minimize this by reallocating finished channels to running chunks as discussed in Section~\ref{sec:mc}, ProMC still can achieve up to 10\% higher throughput than MC for small file dominated dataset.

\begin{figure}
\begin{centering}
 \includegraphics[keepaspectratio=true,angle=0,width=60mm]{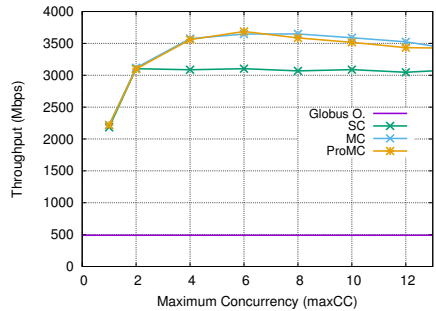}
\caption{Performance comparison of algorithms with the mixed dataset in local-area network.} \label{fig:mixed_openstack}
\end{centering}
\end{figure}

Finally, we conducted experiments in local-area network using the mixed dataset. We setup Globus Connect Personal on the local servers in order to test Globus Online performance. Again, checksum feature is disabled. Globus Online achieved 500 Mbps throughput while other algorithms achieved over 2 Gbps at the worst case. This big difference can be related to the heavy network operations of Globus Connect,  where the personal servers communicate with a central server over the internet in order to operate.

{

\section{Conclusions and Future Work}
\label{sec:conclusion}
We have presented three dynamic algorithms for application-level tuning 
of different protocol parameters for maximizing transfer throughput in wide-area networks.
The parameters tuned by our algorithms (parallelism, pipelining,
and concurrency levels) have shown to be very effective in determining
the ultimate throughput and network utilization obtained by many
data transfer applications.
Though determining the best combination for these parameter values is not a
trivial task, we have shown that our algorithms can choose the parameter combination which
yield demonstrably higher throughputs than baseline and state-of-the-art solutions.

Our algorithms were designed to be client-side techniques and operate 
entirely in user space, and thus special configurations at the server side or at the kernel level
are not necessary to take advantage of them.
The algorithms can be implemented as standalone transfer clients or as
part of an optimization library or service.

In future work, we are planning to include other network and end-system
properties into our formulations, such as disk I/O speed, striped disk
availability, multi-node and multi-path configurations, and sensed network
utilization.
We may also employ other techniques to improve throughput and fairness (such as
block pre-caching, server-side file aggregation, TCP buffer size
clamping) or widen the scope of our algorithms to include additional
transfer parameters depending on the nature of the protocols we wish to
optimize.

\bibliographystyle{elsarticle-num}
\bibliography{references}

\vspace{10mm}
\begingroup
\setlength\intextsep{0pt}
\begin{wrapfigure}{l}{25mm} 
    \includegraphics[width=1in,height=1.25in,clip,keepaspectratio]{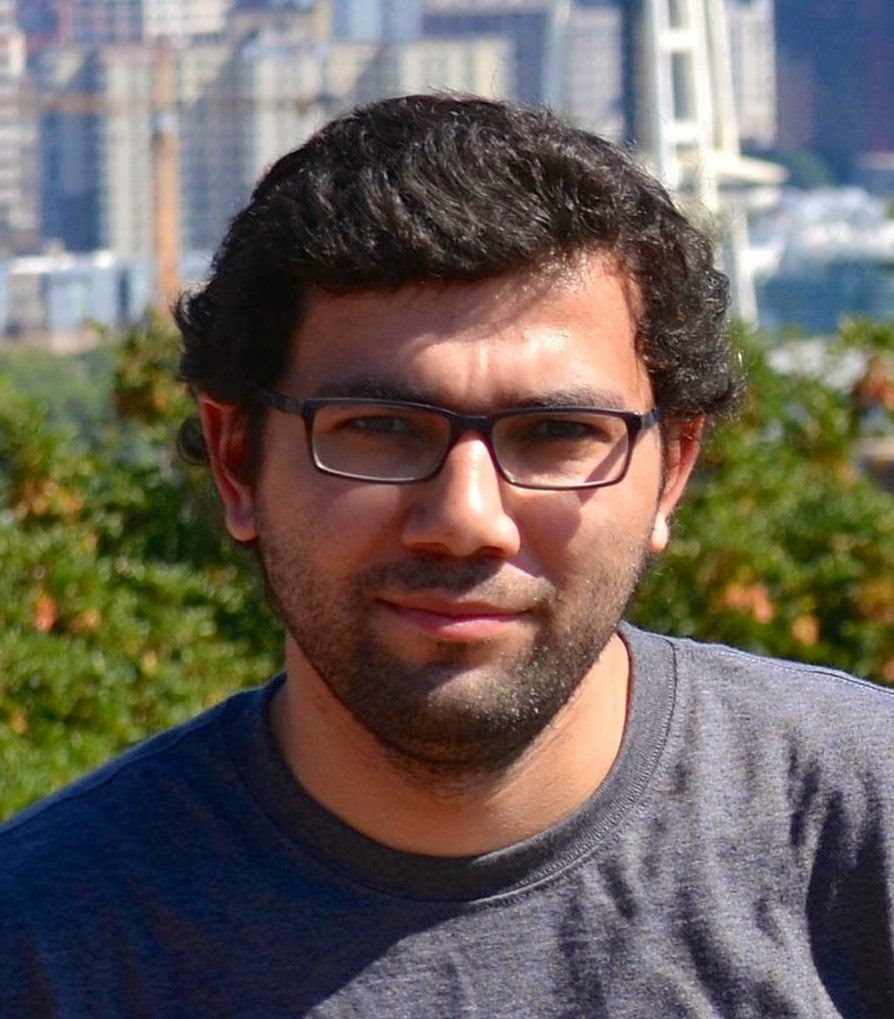}
  \end{wrapfigure}\par
  \textbf{Engin Arslan} is an Assistant Professor at University of Nevada, Reno. He received his BS degree of Computer Engineering from Bogazici University, MS degree from University at Nevada, Reno and PhD degree from Computer Science and Engineering at University at Buffalo, SUNY. He spent one year at National Center for Supercomputing Applications (NCSA) as Postdoctoral Research Associate. His research interests include data intensive distributed computing, high performance computing, distributed systems, and cloud computing.\par
  \endgroup

\begingroup
\setlength\intextsep{0pt}
\begin{wrapfigure}{l}{25mm} 
    \includegraphics[width=1in,height=1.25in,clip,keepaspectratio]{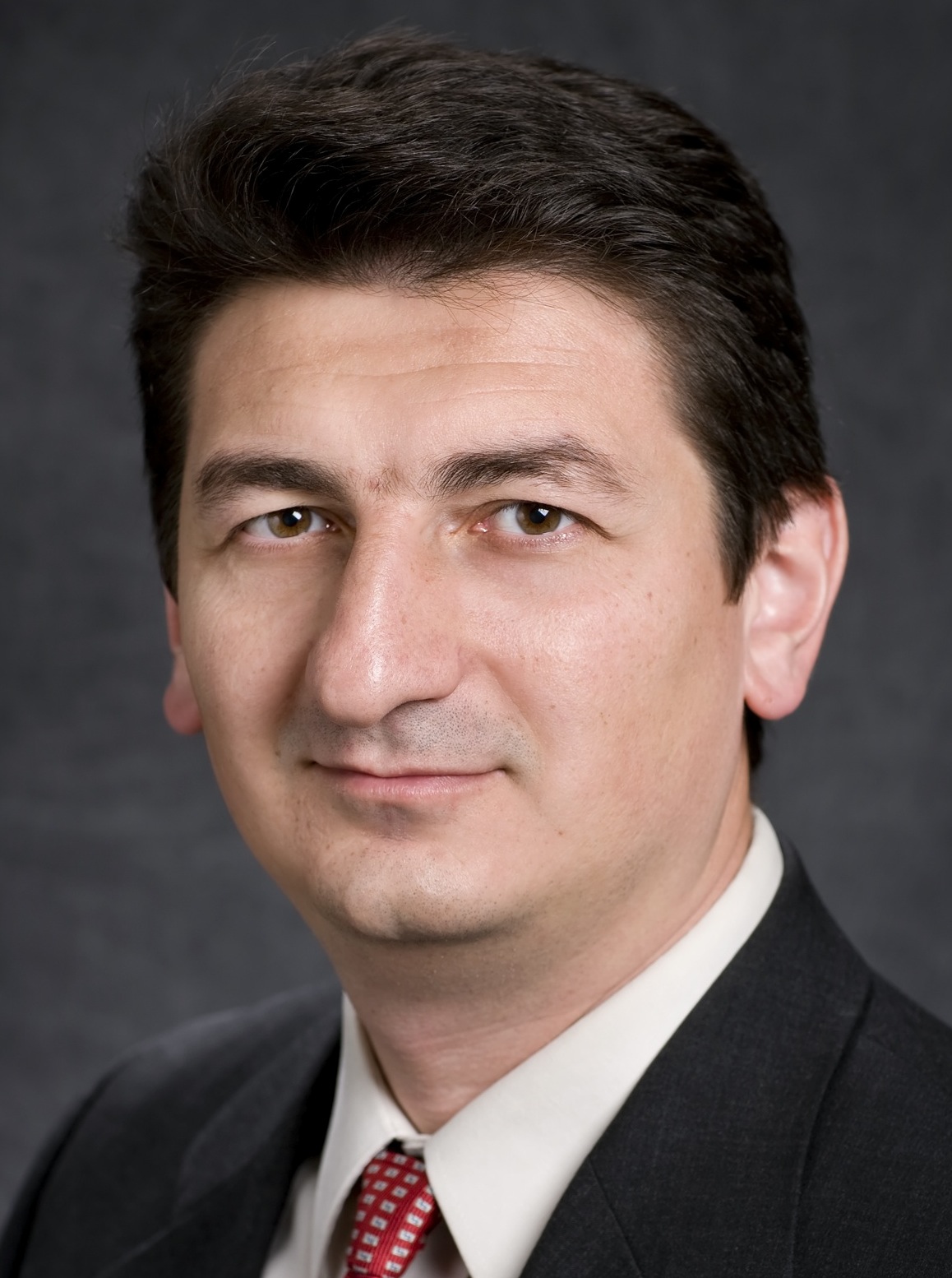}
  \end{wrapfigure}\par
  \textbf{Tevfik Kosar } is an Associate Professor in the Department
of Computer Science and Engineering, University at
Buffalo. Prior to joining UB, Kosar was with the Center for
Computation and Technology (CCT) and the Department
of Computer Science at Louisiana State University. He
holds a B.S. degree in Computer Engineering from
Bogazici University, Istanbul, Turkey and an M.S. degree in
Computer Science from Rensselaer Polytechnic Institute,
Troy, NY. Dr. Kosar has received his Ph.D. in Computer
Science from the University of Wisconsin-Madison. Dr.
Kosar's main research interests lie in the cross-section of
petascale distributed systems, eScience, Grids, Clouds, and collaborative computing
with a focus on large-scale data-intensive distributed applications. He is the
primary designer and developer of the Stork distributed data scheduling system,
and the lead investigator of the state-wide PetaShare distributed storage network
in Louisiana. Some of the awards received by Dr. Kosar include NSF CAREER Award,
LSU Rainmaker Award, LSU Flagship Faculty Award, Baton Rouge Business Report's
Top 40 Under 40 Award, and 1012 Corridor's Young Scientist Award.\par
  \endgroup

\end{document}